\documentclass[pra,aps,preprint,showpacs]{revtex4-2}
\usepackage{amssymb}
\usepackage{amsmath}
\usepackage{amsfonts}
\usepackage{graphicx}
\usepackage{color}
\usepackage{bm}
\usepackage[sort&compress]{natbib}

\renewcommand{\Re}{\operatorname{Re}}
\renewcommand{\Im}{\operatorname{Im}}

\begin{document}

\title{Semiclassical strong-field theory of phase delays in $\omega -2\omega$ above-threshold ionization}
\author{Diego G. Arb\'{o}$^{1,2}$, Sebasti\'{a}n D. L\'{o}pez$^{1},$ and
Joachim Burgd\"{o}rfer$^{3}$}
\affiliation{$^{1}$ Institute for Astronomy and Space Physics - IAFE (CONICET-UBA), CC
67, Suc. 28, C1428ZAA, Buenos Aires, Argentina}
\affiliation{$^{2}$Universidad de Buenos Aires, Facultad de Ciencias Exactas y Naturales
y Ciclo B\'{a}sico Com\'{u}n, Buenos Aires, Argentina}
\affiliation{$^{3}$Institute for Theoretical Physics, Vienna University of Technology,
Wiedner Hauptstra\ss e 8-10/E136, A-1040 Vienna, Austria, EU}
\date{\today }

\begin{abstract}
Phase and time delays of atomic above-threshold ionization were recently 
experimentally explored in an $\omega -2\omega$ setting [Zipp \textit{et al}, Optica \textbf{1}, 361 (2014)].
The phases of wavepackets ejected from argon by a strong $2\omega$ pulse
were probed as a function of the relative phase of a weaker $\omega$ probe pulse.
Numerical simulations solving the time-dependent Schr\"{o}dinger equation (TDSE) displayed a sensitive
dependence of the doubly differential momentum distribution on the relative phase between the $\omega$
and $2\omega$ fields. Moreover, a
surprisingly strong variation of the extracted phase delays on the intensity of the probe pulse was found.
We present a semiclassical strong-field description of the phase delays in the emission of electrons 
in an $\omega -2\omega$ setting and apply it to atomic hydrogen.
Non-perturbative effects in both the $2\omega$ pump and the $\omega$ probe field are included.
The semiclassical description allows tracing phase delays to path interferences between emission
during different points in time of emission within the temporal unit cell of the two-color laser field.
We find good agreement between the semiclassical saddle-point approximation, the full strong field approximation (SFA), and previous results applicable in the perturbative limit of probe fields.
We show that the RABBIT-like perturbative description of phase delays breaks down
for stronger fields and higher-energy electron emission.
In this regime, characterization of the ionization signal requires an entire
ensemble of phase delays {$\delta_i(E)$} with $i=1,2,\ldots$ the difference in photon numbers of the strong
$2\omega$ field involved in the interfering paths. Comparison between SFA and TDSE calculations reveals the influence of the Coulomb field even in this strong-field scenario.
\end{abstract}

\date{\today }
\maketitle

\section{Introduction}

Attosecond chronoscopy of atomic processes has become accessible through pump-probe techniques such as attosecond streaking \cite{Itatani02,
Goulielmakis04,Goulielmakis07} and RABBIT (reconstruction of attosecond
harmonic beating by interference of two-photon transitions) \cite%
{Veniard95,Veniard96,Paul01}. These techniques typically involve lasers of
two very different frequencies: electrons emitted by the absorption
of an XUV photon are probed by a laser pulse in the near-infrared region of the electromagnetic spectrum.
The high sensitivity of the emission spectrum on the relative phase between
the two well-controlled fields has enabled the measurement of the timing of electronic processes on the attosecond scale, in particular of the time zero of the photoionization, and corresponding Eisenbud-Wigner-Smith (EWS) time delays in atoms \cite{Schultze10,Klunder11,Guenot12,Guenot14,Fuchs20}, molecules \cite{Huppert16,Ning14}, and surfaces \cite{Cavalieri07,Neppl12,Ossiander18}.
In attosecond streaking, the oscillating probing field causes classically periodic gains and losses of the final kinetic energy of the emitted electron thereby mapping time information onto energy \cite{Itatani02, Fuchs20}.
In RABBIT, ionization by two consecutive high harmonics of the XUV pulse followed by absorption and emission of a photon of the fundamental frequency opens up two interfering paths to the same final state in the continuum. The relative phase between these two ionization paths permits the interferometric extraction of phase shifts and time delays from the asymptotic behavior of the scattered electron wave packet. Accurate phase and time information can be directly extracted from numerically exact solutions of the time-dependent Schr\"{o}dinger equation (TDSE) \cite{Nagele12,Pazourek12,Kheifets13,Feist14, Su13,Pazourek15,Boll16}.

The concept of measuring phase-shifts and time delays in a RABBIT-like setting was recently extended by Zipp \textit{et al} \cite{Zipp14} to the strong-field regime of two commensurate
frequencies $\omega$ and $2\omega$ (with $\omega$ in the near infrared).
The strong $2\omega$ field induces above threshold ionization (ATI) peaks for argon atoms while the weaker $\omega$ field delayed by a relative phase $\phi$ couples two adjacent ATI peaks at $E_{2n}$ (absorption of $n_{2\omega}$ photons of frequency $2 \omega$) and $E_{2n+2}$ (absorption of $n_{2\omega}+1$ photons of frequency $2 \omega$) to (sideband) states with the same final energy $E_{2n+1}$.
This opens interfering paths (e.g. absorption of $n_{2\omega}$ photons of frequency $2 \omega$
plus one additional $\omega$ photon interfering with absorption of $n_{2\omega}+1$ photons of frequency 
$2 \omega$ plus emission of one $\omega$ photon).
Our recent theoretical study of these $\omega -2\omega$ phase delays employing \textit{ab initio} solution of the TDSE and lowest perturbation theory for atomic argon \cite{Lopez21} revealed surprisingly strong non-linear effects by the probe field beyond lowest-order perturbation theory. This observation raised conceptual questions as to the extraction and interpretation of atomic phase and timing information within such setting in the strong field regime.

In this work, we begin to address some of these questions by exploring
a semiclassical strong-field description of the photoelectron emission from atomic hydrogen in a two-color ($\omega-2\omega$) linearly polarized laser pulse in the multiphoton regime.
The present study involving the saddle-point approximation (SPA) to the strong-field approximation (SFA) represents a generalization of the theory for diffraction at a time grating
\cite{Arbo10a,Arbo10b,Arbo12} to the multiphoton strong-field ionization process.
Key is the observation that the phase-controlled superposition of the $\omega$ and the $2\omega$ fields results in a phase-dependent modulation of the electron emission that gives access to an
entire ensemble of phase delays $\left\{\delta_i(E)\right\}$ with $i=1,2,\ldots$.
The latter are associated with a
plethora of different interfering pathways of multi-photon absorption and emission reaching the same final
state in the continuum. Semiclassically, these interfering pathways can be related to electronic wavepackets emitted at different ionization times within one optical cycle $T=2\pi/\omega$. 
We gauge the applicability of the SPA by comparison with full SFA calculations as well as TDSE results.
In the limit of a very weak $\omega$ probe field the present SPA converges to previous SFA perturbative results for $\delta_{i=1}$ \cite{Zipp14,Lopez21,Bertolino21}.
Within the SPA or SFA, the influence of the Coulomb potential of the outgoing wavepacket is neglected \cite{Faria99,Arbo10b,Lai15,Faria20,Maxwell20}.
The comparison with TDSE results demonstrates the influence of Coulomb effects in the strong-field regime of the present multiphoton strong-field iinterference (MPSFI) scenario.

The structure of the paper is as follows. In Sec. \ref{theory} we briefly review the semiclassical theory of photoionization. We present the generalization of the SPA to the $\omega - 2\omega$ field in the non-perturbative regime in Sec. \ref{SC-delays}, where approximate analytic expressions for the interfering phases between different emission points on the time grating are given.
Results for signatures of these interferences on the doubly differential momentum distributions (DDMD) of emitted electrons as a function of the relative phase $\phi$ between the $\omega$ and the $2\omega$ fields for the atomic phase delays are discussed and a comparison with full SFA and TDSE calculations is given in Sec. \ref{comparison}. By means of a Fourier analysis of angle-resolved energy spectra, the set of phase delays $\left\{\delta_i(E)\right\}$ is extracted for the non-linear regime in Sec. \ref{delays}. Our conclusions are summarized in Sec. \ref{conclusions}. Atomic units ($e=\hbar =m_{e}=1$ a.u.) are used throughout unless stated otherwise.

\section{Brief review of the semiclassical saddle-point approximation}

\label{theory}

Ionization of atomic and molecular systems by a strong laser
pulse is frequently treated in the single-active-electron approximation.
The TDSE for an atom exposed to the laser field reads 
\begin{equation}
i\frac{\partial }{\partial t}\left\vert \psi (t)\right\rangle =\left[
H_{0}+H_{\text{int}}(t)\right] \left\vert \psi (t)\right\rangle ,
\label{TDSE}
\end{equation}%
where $H_{0}=\vec{p}^{2}/2+V(r)$ is the time-independent atomic Hamiltonian,
whose first term corresponds to the electron kinetic energy and its second
term to the electron-core Coulomb interaction. In Eq. (\ref{TDSE}), $H_{%
\text{int}}(t)$ represents the interaction Hamiltonian between the
atomic system and the external radiation field. In the case of hydrogen for which we present numerical results in the following, Eq. (\ref{TDSE}) with $V(r)=-Z/r$, ($Z=1$) is exact. In the presence of the external field, the electron initially bound in an atomic state $|\varphi_{i}\rangle $ can undergo a transition to a final continuum state $|\varphi_{f}\rangle $ with final momentum $\vec{k}$ and energy $E=k^{2}/2$, corresponding to photoionization.
The transition amplitude can be expressed within the time-dependent distorted wave theory in the prior form as \cite{Macri03, Arbo08a}%
\begin{equation}
T_{\mathrm{if}}=-i\int_{-\infty }^{+\infty }dt\,\langle \chi _{f}^{-}(\vec{r}%
,t)|H_{\text{int}}(\vec{r},t)|\varphi _{i}(\vec{r},t)\rangle ,  \label{Tif}
\end{equation}%
where $\varphi _{i}(\vec{r},t)=\varphi _{i}(\vec{r})\,e^{iI_{p}t}$ is the
initial atomic state with ionization potential $I_{p}$ and $\chi _{f}^{-}(%
\vec{r},t)$ is the distorted final state. Eq. (\ref{Tif}) would be still exact provided 
the exact exit channel function $\chi _{f}^{-}(\vec{r},t)$ is used.

Eq. (\ref{Tif}) serves as a starting point for several frequently used approximations. One of the most popular is the SFA, which neglects the Coulomb interaction between the ionized electron and the ionic core in the exit channel. The underlying assumption is that the strong laser field - electron interaction dominates over the Coulomb field. Thus, in the SFA the exact exit channel function $\chi _{f}^{-}(\vec{r},t)$ is reduced to a Volkov state \cite{Volkov35}, i.e., $\chi_{f}^{-}(\vec{r},t)=\chi_{f}^{V}(\vec{r},t)$, where
\begin{eqnarray}
\chi _{f}^{V}(\vec{r},t) &=&\frac{1}{(2\pi )^{3/2}}\exp \{i[\vec{k}+\vec{A}%
(t)]\cdot \vec{r}\}  \notag \\
&\times &\exp \left\{ \frac{i}{2}\int_{t}^{\infty }[\vec{k}+\vec{A}%
(t^{\prime })]^{2}dt^{\prime }\right\}.
\label{Volkov}
\end{eqnarray}
The vector potential $\vec{A}(t)$ is given in terms of the external electric field by $\vec{A}(t)=-\int_{-\infty }^{t}dt^{\prime}\vec{F}(t^{\prime })$.
The Volkov state [Eq. (\ref{Volkov})] represents the solution of the TDSE in the length gauge
for a free electron exposed to an electromagnetic field.
We evaluate the transition matrix element within the dipole approximation by inserting Eq. (\ref{Volkov}) and $H_{\rm{int}}(\vec{r},t)=\vec{F}(t) \cdot \vec{r}$ into Eq. (\ref{Tif}), which yields
\begin{equation}
T_{\mathrm{if}}=\int_{-\infty }^{+\infty }\,M_{\mathrm{if}} (t)\ e^{iS(t)}\,\,dt,
\label{Tm}
\end{equation}
where the coupling matrix element
\begin{equation}
M_{\mathrm{if}} (t) =-i\vec{F}(t)\cdot \vec{d}\left[ \vec{k}+\vec{A}(t)\right]
\label{Mif}
\end{equation}
contains the dipole transition moment defined as $\vec{d}(\vec{v})=(2\pi
)^{-3/2}\langle e^{i\vec{v}\cdot \vec{r}}|\vec{r}|\varphi _{i}(\vec{r}%
)\rangle $, and the phase in Eq. (\ref{Tm}) is given by the Volkov action \cite{Volkov35}
\begin{equation}
S(t) =-\int_{t}^{\infty }dt^{\prime }\left\{ \frac{\left[ \vec{k}+\vec{A}%
(t^{\prime })\right] ^{2}}{2}+I_{p}\right\}.
\label{action}
\end{equation}

Throughout this paper, we consider linearly polarized laser fields (in the $\hat{z}$ direction) featuring a smooth envelope with a central flat-top region spanning $N \gg 1$ optical cycles within which both $\vec{F}(t)$ and $\vec{A}(t)$ are strictly periodic with period $T$.
 From Eq. (\ref{action}) it follows that in the central region the action $S(t)$ satisfies the condition of Floquet periodicity
\begin{equation}
S(t+jT)=S(t)+\widetilde{E}jT,  \label{S-periodic}
\end{equation}
with $j=1,2,\ldots$, which implies, up to a linear shift with time, $T$--periodicity. 
The constant of proportionality for the linear shift of the action corresponds to the (quasi)energy
\begin{equation}
\widetilde{E}=\frac{k^{2}}{2}+I_{p}+U_{p},
\label{a}
\end{equation}
which includes the ponderomotive energy $U_{p}=\int_{t}^{t+T}dt^{\prime} A(t^{\prime})^{2}/2$.
Using the $T$-periodicity of the coupling matrix element $M_{\mathrm{if}}(t+jT)=M_{\mathrm{if}}(t)$,
resulting from the periodicity of the electric field and
vector potential, the transition matrix $T_{\mathrm{if}}$ in Eq. (\ref{Tm}) can be written in terms of the contribution stemming from one cycle or, equivalently, one temporal unit cell 
of the ``time grating'' \cite{Arbo12,DellaPicca20} as
\begin{eqnarray}
T_{\mathrm{if}} &=&\int_{0}^{NT}\,M_{\mathrm{if}} (t)e^{iS(t)}\,\,dt  \notag \\
&=&\sum_{j=0}^{N-1}\int_{jT}^{(j+1)T}M_{\mathrm{if}}(t+jT)e^{iS(t+jT)}dt  \notag \\
&=&\sum_{j=0}^{N-1}e^{i\widetilde{E}jT}\int_{0}^{T}M_{\mathrm{if}} (t)e^{iS(t)}dt  \notag \\
&=&\frac{\sin {(\widetilde{E}TN/2)}}{\sin {(\widetilde{E}T/2)}}\,e^{(i\widetilde{E}T(2N-1)/2)}I_{\mathrm{if}}(\vec{k}),
\label{Tm3}
\end{eqnarray}
with $N$ the total number of optical cycles in the flat-top region.
For simplicity and in order to arrive at analytic results, the contributions from the ramps on and off 
of the field to the total ionization amplitude have been omitted in the current SPA analysis. 
Note, however, that in the full numerical implementations of the SFA and the TDSE presented below
ramp-on and ramp-off effects will be fully included. In Eq. (\ref{Tm3})
\begin{equation}
I_{\mathrm{if}}(\vec{k})=\int_{0}^{T} M_{\mathrm{if}} (t)e^{iS(t)}dt
\label{intra-I}
\end{equation}
denotes the contribution to the transition amplitude stemming from one single optical cycle. 

The DDMD of emitted electrons as a function of the transverse ($k_{\bot}$) and longitudinal ($k_z$) momenta or, equivalently, of the energy ($E$) and angle ($\theta$) with respect to the polarization direction can be expressed in terms of $T_{\mathrm{if}}$ [Eq. (\ref{Tm3})] as
\begin{equation}
P(k_{\bot },k_{z}) = \frac{1}{\sqrt{2E}} P(E,\cos \theta) = 2\pi \left\vert T_{\mathrm{if}}\right\vert ^{2}. 
\label{dPdE}
\end{equation}
Thus, the DDMD can be represented by a product of an intracycle factor
\begin{equation}
F(\vec{k}) = \left|I_{\mathrm{if}}(\vec{k})\right|^2 ,
\label{intra-factor}
\end{equation}
which stems from the contributions within one optical cycle and plays the role of a form factor of the time grating \cite{Arbo08a,DellaPicca20,Arbo08b}, and of the intercycle factor
\begin{equation}
B(\widetilde{E}) = \left( \frac{\sin {(\widetilde{E}TN/2)}}{\sin {(\widetilde{E}T/2)}} \right)^2 ,
\label{inter-factor}
\end{equation}
which corresponds to the Bragg factor of the time grating signifying the 
superposition of contributions from $N$ different optical cycles. The expression for the DDMS can thus
be expressed as diffraction at a time grating \cite{Arbo10a,Arbo10b,Arbo12}, i.e.,
\begin{equation}
P(k_{\bot},k_z) \propto B(\widetilde{E}) F(\vec{k}).
\label{factorization}
\end{equation}
We emphasize that, apart from neglecting transient turn-on and turn-off effects, the decomposition of the DDMS into intracycle and intercycle factors holds in general, in particular also in the present case of temporal modulations of the time grating resulting from the superposition of the $\omega$ and $2\omega$ laser fields.

Finite maxima are reached at the zeros of the denominator of the intercycle
factor $B(\widetilde{E})$ in Eq. (\ref{inter-factor}), i.e., at energies satisfying $\widetilde{E}T/2=n\pi$. Such maxima are recognized as the multiphoton peaks of the photoelectron spectra. They occur at final electron kinetic energies $k^2/2$ equal to
\begin{equation}
E_{n}=n \frac{2 \pi}{T} - I_{p}-U_{p},  
\label{ATI1}
\end{equation}
where we have used Eq. (\ref{a}). In fact, for long pulses when $N\rightarrow \infty $, the intercycle factor becomes a series of delta functions, i.e., $\sum_{n}\delta(E-E_{n})$, expressing the conservation of quasi-energy for multiphoton absorption. Correspondingly, for finite pulse durations $\tau =NT$, each multiphoton peak has a width $\Delta E\sim 2\pi /NT$, consistent with time-energy uncertainty.

The intracycle amplitude [Eq. (\ref{intra-I})] can be numerically evaluated to yield the SFA. In the semiclassical limit, assuming the action $S(t)$ to be large and rapidly varying, Eq. (\ref{intra-I}) can be evaluated within the SPA. Hence, $I_{\mathrm{if}}(\vec{k})$ reduces to a coherent superposition of amplitudes associated with electron bursts emitted with momentum $\vec{k}$ at different times $t_{\beta}$ within a single optical cycle (or unit cell of the temporal lattice) \cite{Gribakin97}
\begin{equation}
I_{\mathrm{if}}(\vec{k})\simeq \sum_{\beta } W(t_{\beta}) e^{iS(t_{\beta })},  
\label{I1}
\end{equation}
with $W(t_{\beta})$ the complex amplitude of the emission burst.
The ionization times $t_{\beta }$ are, in general, complex, fulfilling the saddle-point condition
\cite{Jasarevic20} $\dot{S}(t_{\beta })=0$ (the dot denotes the time derivative), i.e., 
\begin{equation}
\frac{\left[ \vec{k}+\vec{A}(t_{\beta })\right] ^{2}}{2}+I_{p}=0,
\label{saddle}
\end{equation}
where we have used Eq. (\ref{action}). In the appendix we provide details of the saddle-point integration
within the SPA leading to Eq. (\ref{I1}).
Solutions of Eq. (\ref{saddle}) come in pairs ($t_{\beta},t_{\beta }^{\ast }$),
where the star denotes complex conjugation. 
From each pair we select only the exponentially converging (non-diverging) solutions, i.e.,
$ \exp \left\{ -\Im\left[S(t_{\beta })\right] \right\} \leq 1$. 
We refer in the following to the approximation [Eq. (\ref{I1})] with numerically determined $t_{\beta}$
as the full SPA.

\section{Semiclassical approximation for phase delays in $\omega - 2\omega$ ionization}
\label{SC-delays}

We now explore the effect of an $\omega - 2\omega$ laser field on the photoionization dynamics with the goal to relate the phase delays observed with the $\omega - 2\omega$ setting to semiclassical path interferences. The present approach allows for both strong pump ($2\omega$) and probe ($\omega$) fields and is, thus, not restricted to the perturbative limit.

We consider the two-color electric field of the form
\begin{equation}
\vec{F}(t)=f(t)\left[ F_{2\omega}\sin \left( 2\omega t +\phi\right) +F_{\omega}\sin (\omega t )\right] \ \hat{z},
\label{field2}
\end{equation}
with $\phi $ the relative phase of the second harmonic with respect to the
fundamental laser field, $f(t)$ is the normalized envelope function varying between $0$ and $1$,
$\hat{z}$ is the polarization direction of
both fields, and $F_{2\omega }$ and $F_{\omega }$ are the field strengths of
the second harmonic and fundamental frequency, respectively.
For our numerical results we use $F_{2\omega }= 0.05$ a.u. and $F_{\omega} = 0.005$ a.u.
corresponding to Keldysh parameters $\gamma_{2\omega}=2$ and $\gamma_{\omega}=10$ in the multiphoton regime.
The relative phase $\phi$ serves as a control parameter to unravel the subcycle ionization dynamics and ionization. 

\begin{figure}[tbp]
	\includegraphics[width=10cm]{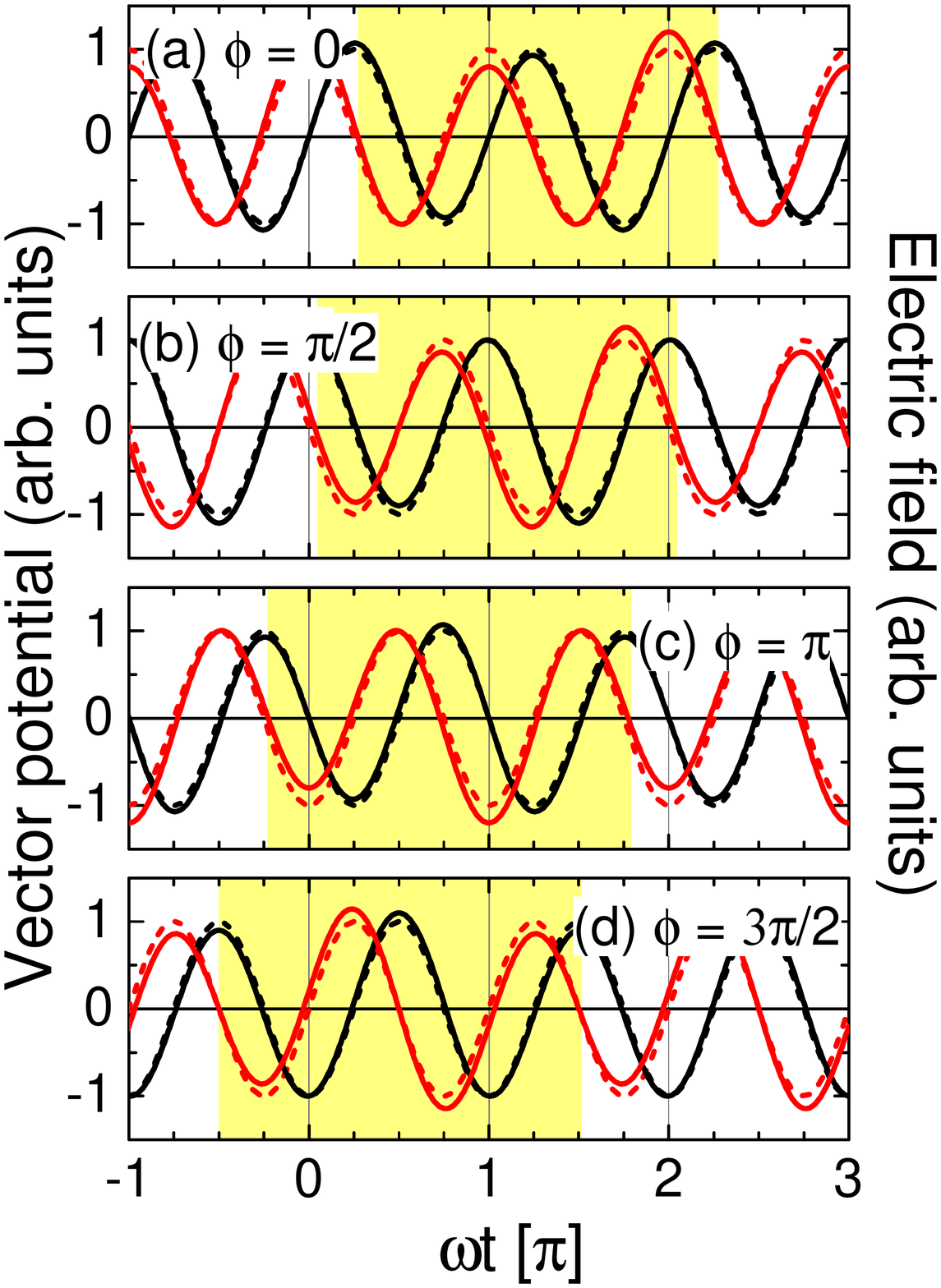}
	\caption{Vector potential (red) and electric field (black) along the polarization axis as a function of time. Solid line: the two-color field [Eqs. (\ref{field2}) and (\ref{Avector})],
dashed line: the $2\omega$ field [$F_{\omega}=0$] in Eqs. (\ref{field2}) and (\ref{Avector})] for reference.
Fields are shown for different relative phases (a) $\phi =0,$ (b) $\phi =\pi/2$,
(c) $\phi =\pi$, and (d) $\phi =3\pi /2$. Yellow shadings delimit the unit cell of the time lattice
determined by zeros of the vector potential.
One-color ($2\omega$) vector potentials and electric fields are scaled to unit amplitude.}
\label{fields-2c}
\end{figure}

For a long pulse with suitable switch-on and switch-off, the vector potential can be written in its central part ($f(t)\simeq 1$), as
\begin{equation}
\vec{A}(t)=f(t)\left[ \frac{F_{2\omega }}{2\omega }\cos (2\omega t +\phi)+\frac{F_{\omega }}{\omega }\cos (\omega t )\right] \hat{z},  \label{Avector}
\end{equation}
displaying the same periodicity as the electric field,
i.e., $\vec{A}(t)=\vec{A}(t+2j\pi /\omega )$ and $\vec{F}(t)=\vec{F}(t+2j\pi /\omega )$,
with $j$ any integer number provided that $f(t+2j\pi /\omega )=1$. Note that the
periodicity of the time grating or ``lattice constant'' of the temporal lattice is determined by the fundamental frequency $\omega$, i.e., $T=2\pi / \omega$ and not by the $2\omega$ field.
The latter, even though stronger than the $\omega$ field ($F_{2\omega } > F_{\omega }$),
should rather be viewed as a strong and rapid modulation with period $T/2$
of the time grating with period $T$.
Fig. \ref{fields-2c} displays the time dependence of the superimposed $\omega - 2\omega$ field.
For reference, the dominant $2\omega$-one-color field ($F_{\omega}=0$) is also given.
The unit cell corresponding to one optical cycle of the field
is delimited by the zeros of the vector potential \cite{Arbo10a,Arbo10b}.
For the limiting one-color case ($F_{\omega} \rightarrow 0$) this would correspond to,
for example for $\phi=0$, $t\in \left[ \pi/4\omega, 9\pi/4\omega \right)$ (Fig. \ref{fields-2c}a),
while these values vary for the two-color case.
Accordingly, the unit cell of the time grating for the two-color pulse for $\phi =0$ is located at 
$t\in \left[ 0.275 \pi/\omega, 2.275\pi /\omega \right) $ (Fig. \ref{fields-2c}a).
Analogous shifts can be found for other values of $\phi$.

\begin{figure}[tbp]
	\includegraphics[width=10cm]{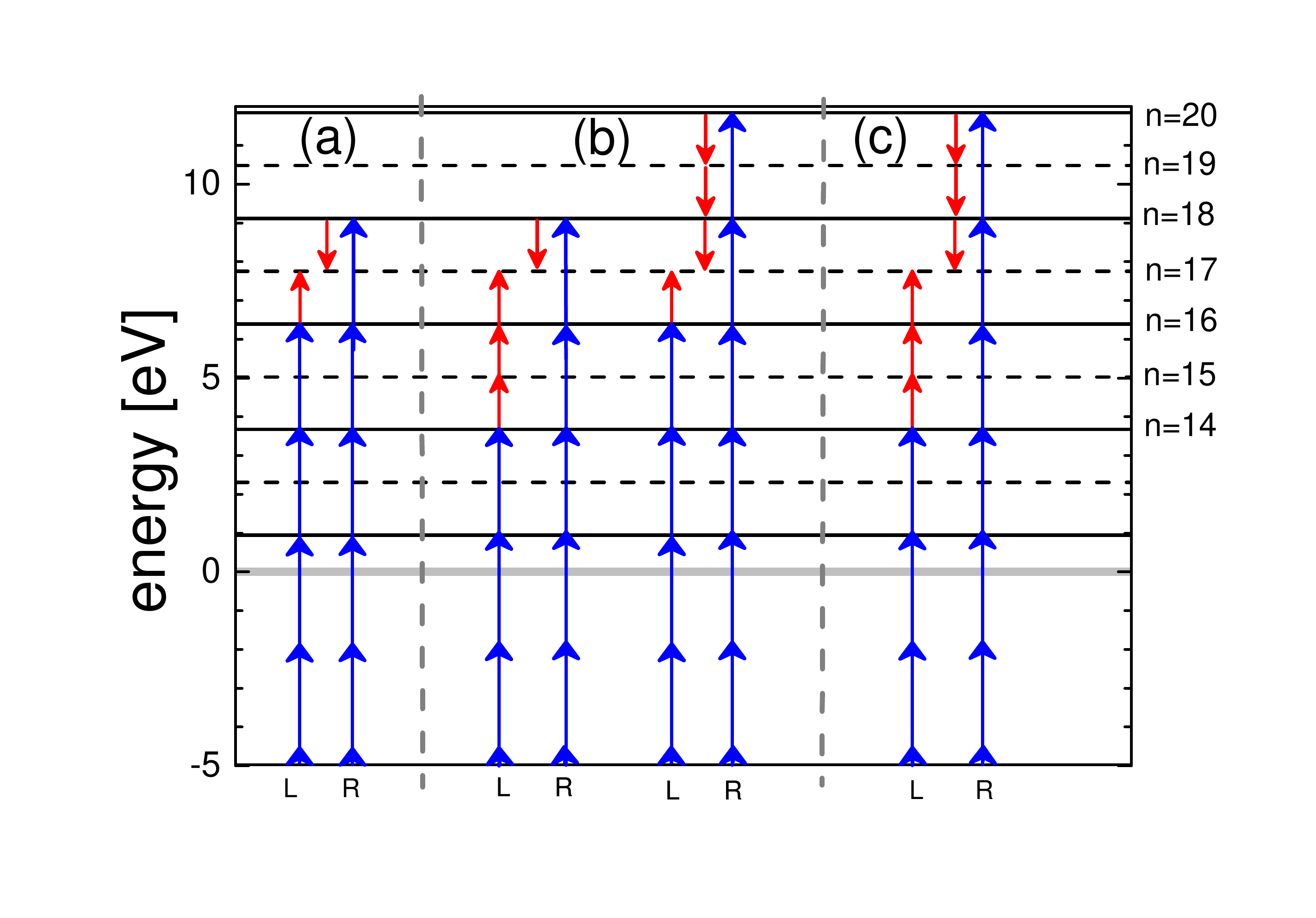}
	\caption{Examples for interfering photon absorption/emission path ways to the same final state,
	the ``sideband'' level ($n=17$). The paths are sorted according to the difference in the number 
	$i = \left| n_{2\omega} - n^{\prime}_{2\omega} \right|$ of the strong-field $2\omega$ photons (blue arrows) involved, (a) $i=1$, (b) $i=2$, and (c) $i=3$. Red arrows denote probe photons with frequency $\omega$.
	 An analogous path classification can be applied to paths leading to ATI peaks.}
\label{paths}
\end{figure}

The dependence of the angular differential ionization probability [Eq. (\ref{dPdE})], 
e.g. in the forward direction  ($\cos \theta = 1$), on the relative phase $\phi$ between the $\omega$ and $2\omega$ fields can be Fourier expanded as \cite{Lopez21,Zhou22}
\begin{equation}
P(E,\cos \theta = 1) = c_0(E) + \sum_{i\geq1} c_i(E) \cos\left[ i\phi-\delta_i(E) \right] .
\label{fourier}
\end{equation}
This expansion signifies the quantum path interference between different multi-photon absorption and emission
pathways to the same final state \cite{Lopez21}. Adopting the standard RABBIT terminology, we refer to peaks in the energy spectra at $E_n$ with $n$ even to ATI peaks and to side bands when $n$ is odd [see Eq. (\ref{ATI1})]. The phase-independent term $i=0$ mirrors the $\phi$-average forward spectrum corresponding to an experiment with randomly fluctuating relative phase $\phi$.
The order of the Fourier component corresponds to the difference 
$i = \left| n_{2\omega} - n^{\prime}_{2\omega} \right|$ in the number ($n_{2\omega},n^{\prime}_{2\omega}$)
of strong-field $2\omega$ photons involved in the two interfering paths (Fig. \ref{paths}).
The first-order Fourier component $i=1$ in Fig. \ref{paths}a bears closest resemblance to the RABBIT protocol
\cite{Veniard95,Veniard96,Paul01,Zipp14,Lopez21} involving the absorption/emission of just one 
($n_{\omega}=1$) probe (weak-field) photon $\omega$.
This Fourier component is expected to dominate the $\phi$ variation of the
DDMD in the perturbative limit of weak probe fields ($F_{\omega}/F_{2\omega} \ll 1$).
With increasing amplitude $F_{\omega}$ of the probe field higher-order Fourier components ($i > 1$) will
account for non-perturbative effects on the pump-probe protocol.
Accordingly, the set of atomic phase shifts $\left\{\delta_i(E)\right\}$, $i=1,2,\ldots$ provides
detailed information on the atomic ionization dynamics in this multi-photon scenario.
For simplicity, we refer in the following to $\delta_i(E)$ as phase delays even though both positive
and negative values of $\delta_i(E)$ are possible.
In the perturbative limit only $i=1$ is present, thus, $\delta_{i=1}(E)$
can be viewed as a finite-difference approximation to the spectral derivative of the scattering phase shift
and, thus, as a time delay. In the presence of higher Fourier components in Eq. (\ref{fourier}), such an
interpretation is no longer obvious.
One goal of the present work is to relate the multi-photon quantum pathway interference encapsulated in
Eq. (\ref{fourier}) (Fig. \ref{paths}) to the semiclassical wavepacket interference in the time domain.
We calculate in the following the DDMD, $P(k_{\bot},k_z)$, and $P(E,\cos \theta)$, [Eq. (\ref{dPdE})]
as well as the atomic phase delays $\delta_j(E)$ [Eq. (\ref{fourier})] for the $\omega-2\omega$ field
in the semiclassical SPA, the SFA, as well as the TDSE.

Under the assumption of an adiabatic switch-on and switch-off, the action $S(t)$ 
[Eq. (\ref{action})] entering the SPA can be analytically calculated 
(in the central region where $f(t)=1 $) as
\begin{eqnarray}
S(t) &=&\widetilde{E}t+b\sin (2\omega t+\phi)+c\sin (4\omega t + 2\phi)+d\sin (\omega t)
\label{action2} \\
&&+e\sin (\omega t+\phi )+f\sin (2\omega t)+g\sin (3\omega t+\phi ), 
\notag
\end{eqnarray}
where $\widetilde{E}$ is given by Eq. (\ref{a}) and
\begin{eqnarray}
b &=&\frac{F_{2\omega }}{4\omega ^{2}}\hat{z}\cdot \vec{k},  \label{abc} \\
c &=&\frac{U_{p,2}}{4\omega },  \notag \\
d &=&\frac{F_{\omega }}{\omega ^{2}}\hat{z}\cdot \vec{k},  \notag \\
e &=&\frac{F_{2\omega }F_{\omega }}{4\omega ^{3}},  \notag \\
f &=&\frac{U_{p,1}}{2\omega },  \notag \\
g &=&\frac{F_{2\omega }F_{\omega }}{12\omega ^{3}}.  \notag
\end{eqnarray}
In Eq. (\ref{abc}) $U_{p}=U_{p,2}+U_{p,1}=F^{2}_{2\omega }/(4\omega )^{2}+F^{2}_{\omega}/(2\omega )^{2}$
is the sum over the ponderomotive energies contributed by each color.
We have omitted diverging terms common to all electron trajectories in Eq. (\ref{action2})
since only differences between actions will be relevant.
The two-color action reduces to the one-color action
when either $F_{\omega }=0$ (for $d=e=f=g=0$)
or $F_{2\omega }=0$ (for $b=c=e=g=0$) \cite{Arbo10a,Arbo10b}.  

Within one temporal unit cell, the laser field features four extrema (Fig. \ref{fields-2c}) 
near each of which an electronic wavepacket can be emitted at times 
$t_{\beta}\;\left(\beta=1,...,4\right)$ [Eq. (\ref{I1})].
In the presence of both the strong $2\omega$ field $~ F_{2\omega}$ and the weaker $\omega$ probe field 
$~ F_{\omega}$, the determination of the ionization times $t_{\beta}$ requires the numerical solution of the coupled equations [Eq. (\ref{saddle})] for the real and imaginary parts of $t_{\beta}$
\begin{subequations}
\begin{eqnarray}
\frac{F_{2\omega }}{2\omega }\cos \left( 2\omega \Re t_{\beta } + \phi\right)
\cosh (2\omega \Im t_{\beta }) \notag \\
+\frac{F_{\omega }}{\omega }\cos \left(
\omega \Re t_{\beta } \right) \cosh \left( \omega \Im %
t_{\beta }\right) &=&-k_{z}  
\label{couple1} \\
\frac{F_{2\omega }}{2\omega }\sin \left( 2\omega \Re t_{\beta } + \phi \right)
\sinh (2\omega \Im t_{\beta })  \notag \\
+\frac{F_{\omega }}{\omega }\sin \left(
\omega \Re t_{\beta } \right) \sinh \left( \omega \Im %
t_{\beta }\right) &=& \pm\sqrt{2I_{p}+k_{\bot}^{2}},
\label{couple2} 
\end{eqnarray}
\end{subequations}
with $\beta =1,2,3,$ and $4$. The $\pm$ sign in Eq. (\ref{couple2}) must be chosen such that 
in the limit $F_{\omega} \rightarrow 0$, Eqs. (\ref{couple1}) and (\ref{couple2}) yield the well-known analytic solutions for the ionization times for a one-color ($2\omega$) field  \cite{Arbo08a,Arbo08b,Arbo10a,Arbo10b}.

The DDMD [Eq. (\ref{dPdE})] is governed by the product [Eq. (\ref{factorization})] of the Bragg factor 
$B(\tilde{E})$ signifying intercycle interference [Eq. (\ref{inter-factor})] and the structure factor $F(\vec{k})$ representing the intracycle interference [Eq. (\ref{intra-factor})]. In particular,
the latter contains the information on the $\omega-2\omega$ interferences.
From Eqs. (\ref{I1}) and (\ref{C}) in the appendix, the intracycle amplitude stemming from the electron
trajectories with release times $t_{\beta }$ ($\beta =1,\ldots, 4$) can be approximated by
\begin{eqnarray}
I_{\mathrm{if}}(\vec{k}) &\simeq & \frac{2 \sqrt{2} (2 I_p)^{5/4}}{F(t_1^0) \sqrt{2 I_p +k_{\bot}^2}} \left[ e^{i \Sigma(t_{1})} + e^{i \Sigma(t_{2})}
+e^{i \Sigma(t_{3})}+e^{i \Sigma(t_{4})}\right]   \label{terms} \\
&\simeq &\frac{ 4 \sqrt{2} (2 I_p)^{5/4}}{F(t_1^0) \sqrt{2 I_p +k_{\bot}^2}} \left[ e^{i\bar{\Sigma}_{1,2}} 
\cos \left( \frac{\Delta  \Sigma_{1,2}}{2} \right) +
e^{i\bar{\Sigma}_{3,4}} \cos \left( \frac{\Delta  \Sigma_{3,4}}{2} \right) \right]   \nonumber
\end{eqnarray}
with
$\overline{ \Sigma}_{\beta,\beta^{\prime}}=\left[  \Sigma(t_{\beta})+ \Sigma(t_{\beta^{\prime}})\right] /2$ is the mean action of the wavepackets emitted at $t_{\beta}$ and $t_{\beta^{\prime}}$,
and $\Delta  \Sigma_{\beta,\beta^{\prime}}= \Sigma(t_{\beta^{\prime}})- \Sigma(t_{\beta})$,
the action difference between $t_{\beta}$ and $t_{\beta^{\prime}}$.
The modified action entering Eq. (\ref{terms}) is defined as $\Sigma(t) = S(t) + \alpha(t)$,
where $\alpha(t)=-\arg \ddot{S}(t)$ [see Eq. (\ref{C}) in the appendix].
In the prefactor of Eq. (\ref{terms}) we have approximated the ionization times by their values
to zeroth order in the $\omega$ field using Eq. (\ref{saddle}).
These zeroth-order ionization times can be analytically determined as
\begin{eqnarray}
t_{1}^0 &=&\frac{1}{2\omega } \left\{ \cos ^{-1}\left[ \frac{2\omega }{F_{2\omega}} %
\left( -k_{z}-i\sqrt{2I_{p}+k_{\bot}^{2}}\right) \right] -\phi \right\},  \notag \\
t_{2}^0 &=&\frac{1}{2\omega }\left\{- \cos ^{-1}\left[ \frac{2\omega }{F_{2\omega}} %
\left( -k_{z}+i\sqrt{2I_{p}+k_{\bot}^{2}}\right) \right] -\phi + 2\pi \right\},  
\label{ts0p}
\end{eqnarray}%
for $k_{z}\geq 0$ and,%
\begin{eqnarray}
t_{1}^0 &=&\frac{1}{2\omega }\left\{- \cos ^{-1}\left[ \frac{2\omega }{F_{2\omega}} %
\left( -k_{z}+i\sqrt{2I_{p}+k_{\bot}^{2}}\right) \right] -\phi + 2\pi \right\},  \notag \\
t_{2}^0 &=&\frac{1}{2\omega }\left\{ \cos ^{-1}\left[ \frac{2\omega }{F_{2\omega}} %
\left( -k_{z}-i\sqrt{2I_{p}+k_{\bot}^{2}}\right) \right] -\phi + 2\pi \right\},  
\label{ts0m}
\end{eqnarray}
for $k_{z}\leq 0$. 
At this level of approximation, the absolute values of the prefactors of each term representing a wavepacket
are identical. We have also used 
$\left|F(t_{\beta}^0)\right|=\left|F(t_{\beta^{\prime}}^0)\right|, (\beta,\beta^{\prime} = 1,\ldots 4)$
in Eq. (\ref{terms}). The action differences $\Delta \Sigma_{1,2}$ and $\Delta \Sigma_{3,4}$
in Eq. (\ref{terms}) control the interference phase of two electron trajectories being born within the same  ($\Delta \Sigma_{1,2}$ for the first, $\Delta \Sigma_{3,4}$ for the second) half-cycle.

In order to pinpoint to the origin of the interference and to arrive at a simple analytic expression
we simplify Eq. (\ref{terms}) further by using the zeroth-order approximation of the action in the probe field,
$\left( \Delta \Sigma_{0}\right) _{1,2}=\left( \Delta \Sigma_{0}\right) _{3,4} \equiv \Delta \Sigma_{0}$,
where $\Delta \Sigma_{0}$ denotes the one-color ($2\omega$) action difference between
trajectories released during the same optical halfcycle. With this additional assumption, the
form factor simplifies to
\begin{eqnarray}
F(\vec{k}) = \left|I_{\mathrm{if}}(\vec{k})\right|^2 &\simeq& D \left\vert e^{i\overline{ \Sigma}_{1,2}}+e^{i\overline{ \Sigma}_{3,4}}\right\vert^{2}
\cos^{2}\left[ \frac{\Delta  \Sigma_{0}}{2} \right]  \notag \\
&\simeq& 4 D \underset{\mathrm{inter-halfcycle}}
{\ \underbrace{\cos^{2}\left(\frac{\Delta  S}{2}\right) }} \underset{\mathrm{intra-halfcycle}}{\underbrace{\cos ^{2}\left[ \frac{\Delta  \Sigma_{0}}{2} \right] }},
\label{interf}
\end{eqnarray}
with $\Delta S=\overline{S}_{3,4}-\overline{S}_{1,2}=\overline{\Sigma}_{3,4}-\overline{\Sigma}_{1,2}$ and
\begin{equation}
D = \frac{8 (2 I_p)^{5/2}}{F_{2\omega}^2 (2 I_p +k_{\bot}^2) \left| 1-\left[\frac{\omega^2}{F_{2\omega}^2}\left(-k_z+i\sqrt{2I_p+k_{\bot}^2}\right)\right]^2 \right| } .
\end{equation}
To arrive at Eq. (\ref{interf}) we have exploited the $T/2$ periodicity of $\Sigma_{0}(t)$ [Eq. (\ref{Mif})],
and used $\ddot{S}(t_{1}^{0})= i F_{2 \omega} \sqrt{2 I_p+k_{\bot}^2/2} 
\left\{ 1-\left[ 2 \omega/F_{2 \omega} (-k_z + i \sqrt{2 I_p+k_{\bot}^2/2}) \right]^2 \right\} $.
We refer in the following to Eq. (\ref{interf}) as the analytic SPA to distinguish this simplified expression which involves several additional approximations
including the zeroth-order approximation to $t_{\beta}$ from the full SPA [Eq. (\ref{I1})] 
within the analytic approximation.
The intracycle factor $\left|I_{\mathrm{if}}(\vec{k})\right|^2$ factorizes now in
(i) the intra-halfcycle interference factor $\cos^{2}\left[ \Delta \Sigma_{0}/2 \right]$
stemming from the interference of the two electron trajectories released
during one half optical cycle of the $\omega $ field 
(or within one optical cycle of the $2\omega $ field) and 
(ii) the inter-halfcycle interference factor $\cos^{2}\left[ \Delta S/2 \right]$
between the contributions stemming from the two different half cycles of the $\omega $ field
(or, equivalently, between two subsequent optical cycles of the $2\omega $ field).
The two half-cycles are separated from each other by a zero of the vector potential and, in general,
do not have (necessarily) the same duration (see Fig. \ref{fields-2c}) since the $\omega - 2\omega$ field breaks the inversion symmetry within the temporal unit cell.

Ionization phases and phase delays can now be deduced from the interferences between the electron trajectories
associated with the different release times $t_{\beta}$ [Eq. (\ref{I1})].
In order to arrive at an approximate analytic expression for the inter-halfcycle interference factor we
insert the zeroth-order approximation to the ionization times 
$t_{\beta}=(2\beta-1) \pi/4\omega - \phi/2\omega \;\left(\beta=1,...,4\right)$,
applicable in the perturbative limit ($F_{\omega} \ll F_{2\omega}$) of the probe field into $\Delta S$,
resulting in
\begin{eqnarray}
\Delta S &=&n\pi - \left(d+e+g \right) \left[ \cos (\phi/2 +\pi/4) + \sin (\phi /2 +\pi/4) \right] 
\label{DS} \\
&=&n\pi -2\chi \cos \left( \phi/2 \right) .  \notag
\end{eqnarray}
The amplitude $\chi$ of the $\phi$ oscillation follows as
\begin{equation}
\chi =\left( d+e+g\right) /\sqrt{2} = F_{\omega }\left( k_{z}/\omega
^{2}+F_{2\omega }/(3\omega ^{3})\right) /\sqrt{2} .
\end{equation}
This amplitude depends on the wavenumber $k_z$ (or energy) of the emitted electron as well as on the product of both field amplitudes ($\propto F_{\omega} F_{2\omega}$) signifying a non-linear pump-probe response.
Evaluating now the inter-halfcycle factor in Eq. (\ref{interf}) to lowest non-vanishing order
in $\chi$ ($\propto \chi^2$) near an ATI peak $E_n$ (with $n$ even) gives
\begin{equation}
\cos ^{2}[\chi \cos \left( \phi/2 \right) ] \simeq 
1-\frac{\chi ^{2}}{2}+\frac{\chi ^{2}}{2}\cos \left(\phi - \pi\right) +O(\chi ^{4}).  \label{PATI}
\end{equation}
Comparison to Eq. (\ref{fourier}) shows that, to this order, the ionization phase delay near an ATI peak is
completely characterized by its first-order Fourier component $\delta_{i=1}$ with $\delta_i=\pi$.
Correspondingly, near a sideband peak (with $n$ odd), the inter-halfcycle factor reads
\begin{equation}
\sin ^{2} \left[ \chi \cos \left( \phi/2 \right) \right] \simeq 
\frac{\chi ^{2}}{2}+\frac{\chi ^{2}}{2}\cos \left(\phi  \right) + O(\chi ^{4}),  \label{PSB}
\end{equation}
and the corresponding phase delay is $\delta_1=0$ \cite{Zipp14,Bertolino21}.
These results within the SPA reproduce the well-known SFA predictions for the perturbative limit \cite{Zipp14,Lopez21,Bertolino21}.
They provide, in addition, insights into their origin in terms of semiclassical path interferences:
the quantum interference between paths involving a different number of strong-field $2\omega$ photons
($i = \left| n_{2\omega} - n^{\prime}_{2\omega} \right| = 1$) [Fig. \ref{paths}a] can be mapped onto
the temporal interference of wavepackets emitted at different half-cycles of the $\omega$ field or,
equivalently, different cycles of the $2\omega$ field. 

\begin{figure}[tbp]
	\includegraphics[width=12cm]{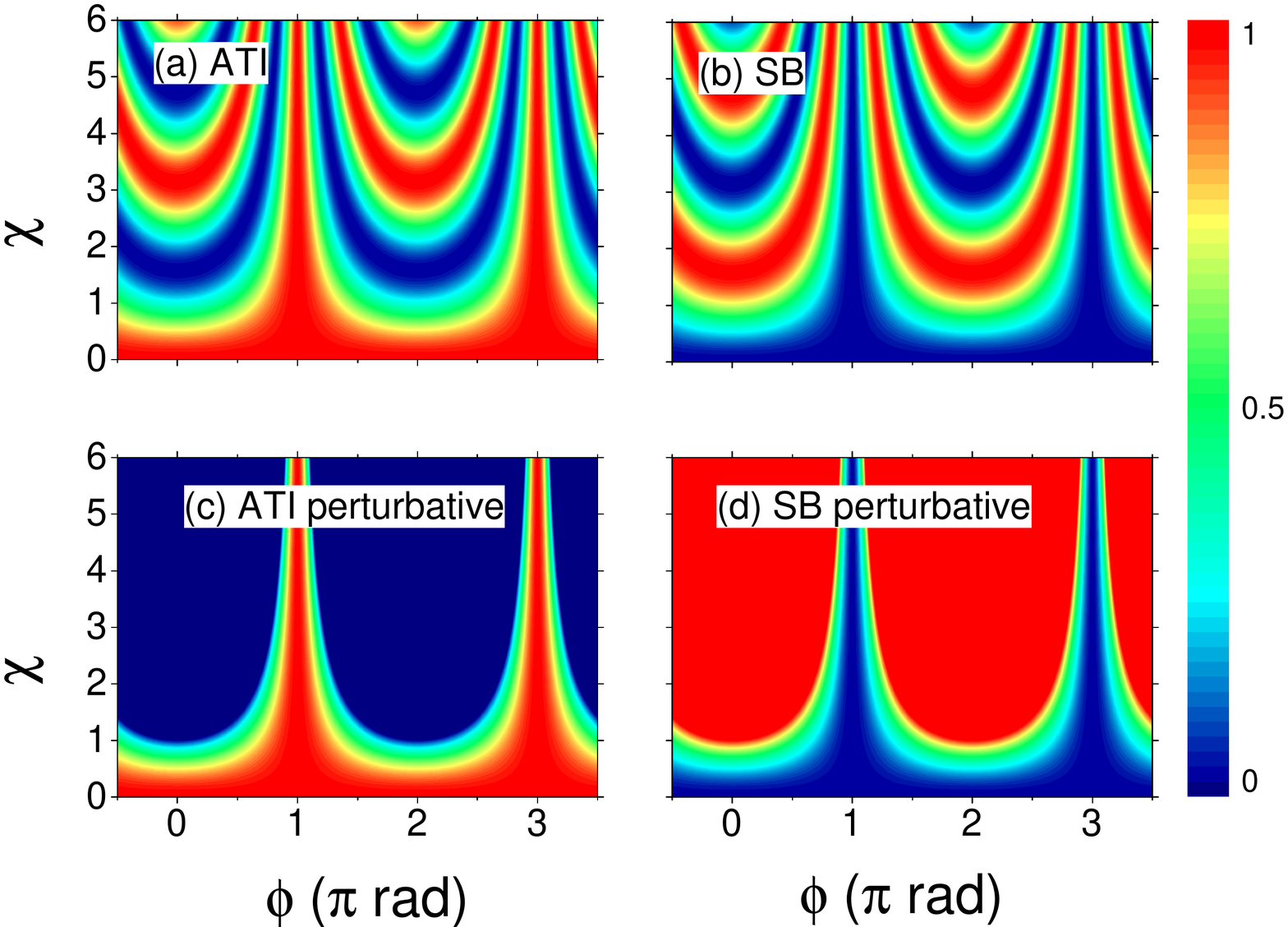}
	\caption{SPA inter-halfcycle factor $\cos^2(\Delta S/2)$ near ATI peaks (a,c) and sidebands (b,d). 
	The factor to all orders is displayed in the top row (a,b), and the perturbative predictions to order $\chi^2$ in the bottom row (c,d).
In the perturbative  limit $\chi \ll 1$ ATI maximizes at $\delta=\pi$ and the sidebands at $\delta=0$.}
\label{analytic-chi}
\end{figure}

\begin{figure}[tbp]
	\includegraphics[width=8cm]{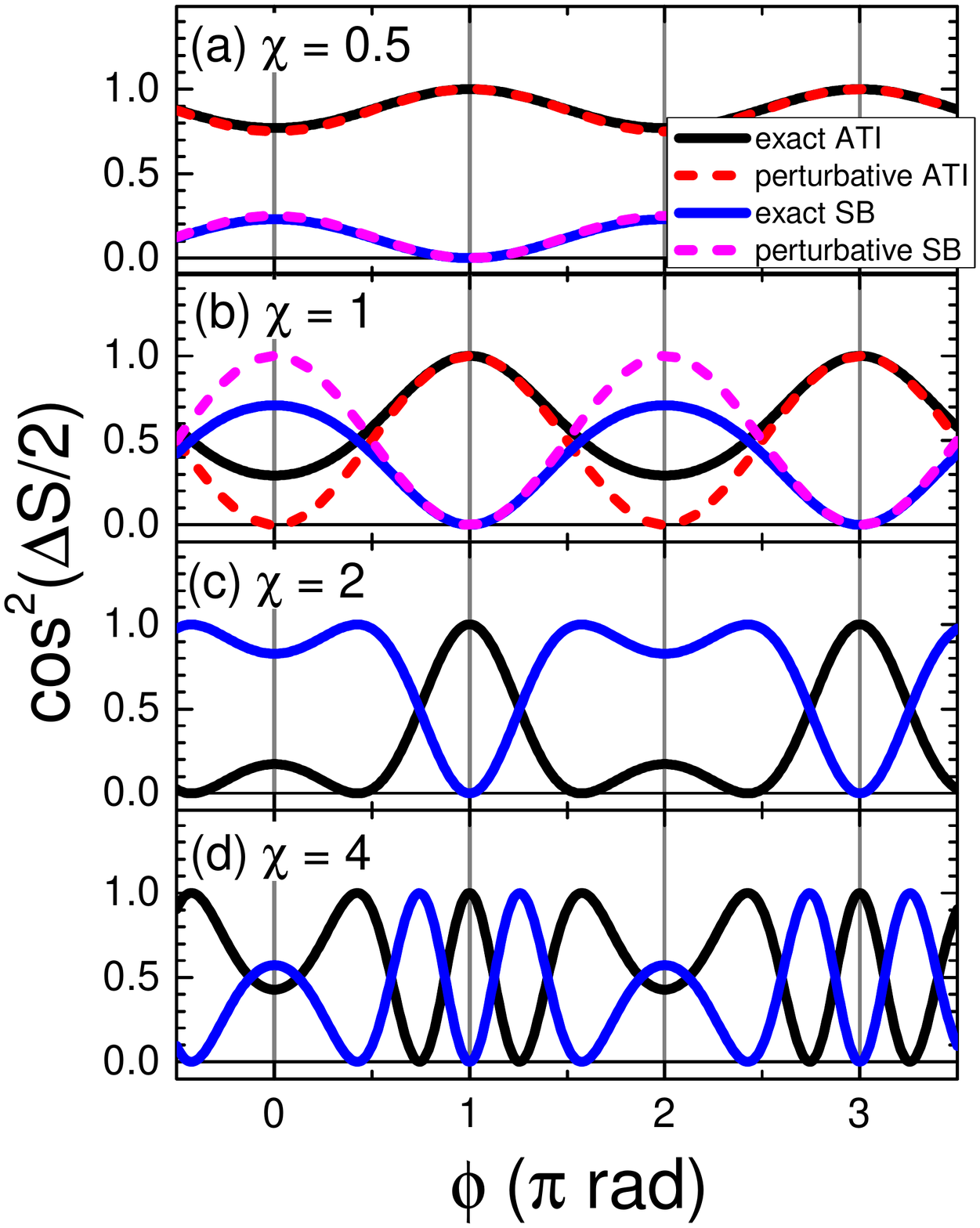}
	\caption{SPA inter-halfcycle interference factor $\cos^2(\Delta S/2)$
	for ATI peaks and sidebands  given by Eqs. (\ref{PATI}) and (\ref{PSB})
	and their respective perturbative predictions as a function of $\phi$ for 
different values of the perturbation parameter: (a) $\chi=0.5$, (b) $\chi=1$, (c) $\chi=2$, and (d) $\chi=4$. In the perturbative case ($\chi=0.5$) the ATI peaks maximize at $\phi=\delta=\pi$ and $3\pi$ while the sidebands have maximum at $\phi=\delta=0$ and $2\pi$. For $\chi > 1$, the perturbative limit fails.}
\label{Fig-anal-approx-phase}
\end{figure}

Going beyond the lowest order in $\chi$, the inter-halfcycle 
interference factor $\cos ^{2}\left[ \Delta S/2 \right]$ as a function of the relative phase
between the $\omega$ and $2\omega$ fields and the strength of $\chi$ drastically varies 
(Figs. \ref{analytic-chi}a,b) andstrongly differs from its perturbative limit (Figs. \ref{analytic-chi}c,d).
Cuts through Fig. \ref{analytic-chi} at fixed $\chi$ are displayed in Fig. \ref{Fig-anal-approx-phase}.
For small $\chi$ a simple sinusoidal variation of the SPA inter-halfcycle interference factor
is observed in line with the RABBIT-like extraction protocol as
the phase shift in the first term of the Fourier expansion [Eq. (\ref{fourier})].
For $\chi=0.5$ (Fig. \ref{Fig-anal-approx-phase}a) the SFA predictions $\delta_1=\pi,3\pi$ for ATI and 
$\delta_1=0,2\pi$ for the sidebands are reproduced by the SPA.
However, for larger $\chi$ (Fig. \ref{Fig-anal-approx-phase}b-d) the present non-perturbative 
SPA results clearly indicate that application of the standard RABBIT-like protocol would fail
as higher-order Fourier
components [Eq. (\ref{fourier})] strongly distort the sinusoidal signal.
Such differences are to be expected since the standard RABBIT protocol is designed for two-photon processes
(one XUV and one IR photon) while the present scenario deals with multiphoton processes involving
many $n_{2\omega}$ and (up to) several $n_{\omega}$ photons.

\begin{figure}[tbp]
	\includegraphics[width=10cm]{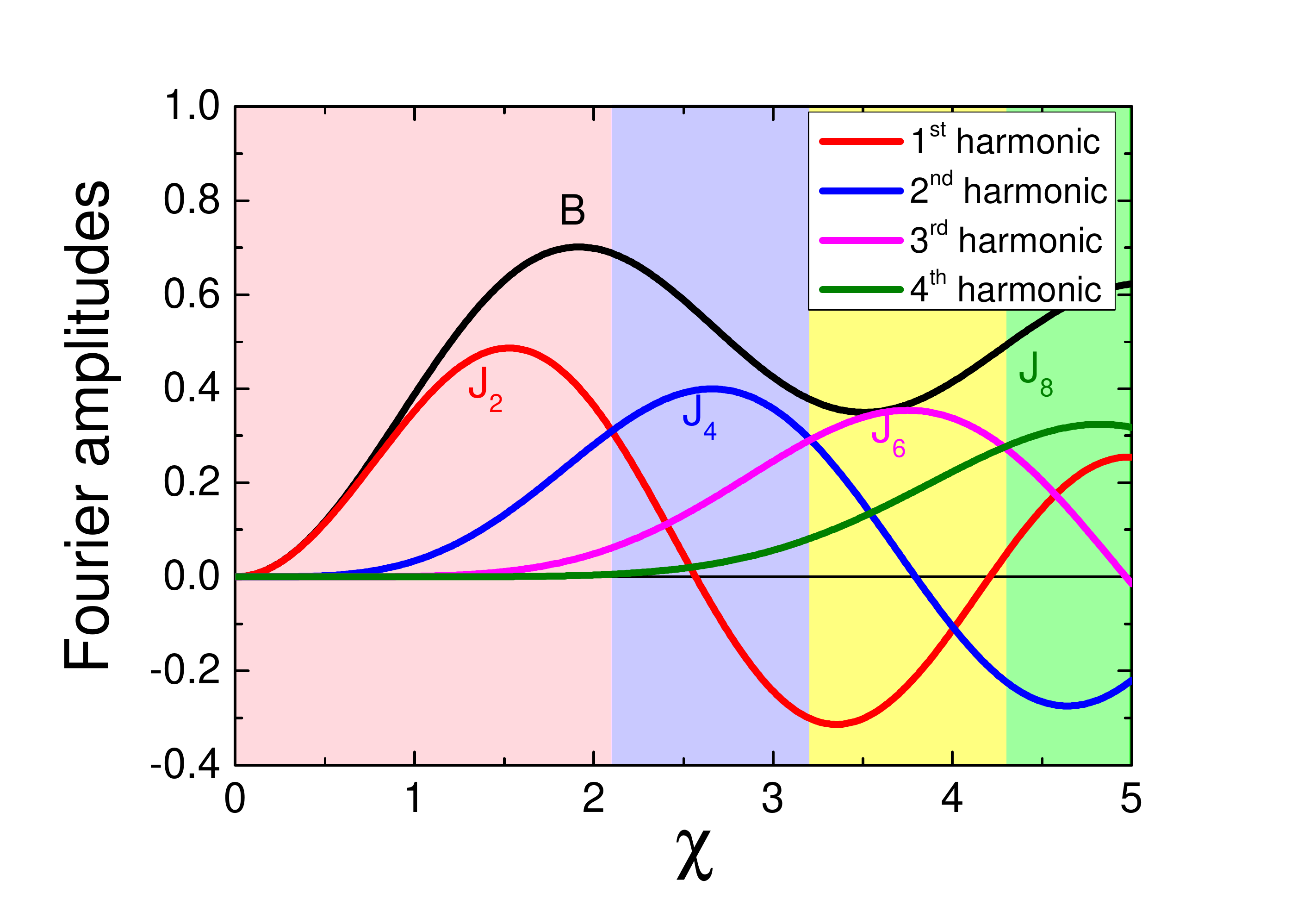}
	\caption{Fourier amplitudes [Eq. (\ref{FSB})] as a function of $\chi$. Whereas the amplitudes of the
	non-oscillatory background (B) and the first harmonic dominate for small $\chi$, the regions with dominant
	higher harmonics for larger $\chi$ are indicated with different color shadings.}
\label{fourier-amplitudes}
\end{figure}

It is therefore convenient to employ an alternative Fourier representation of the inter-halfcycle
interference factors of the analytical SPA valid to all orders in $\chi$,
\begin{equation}
\cos ^{2}[\chi \cos \left( \phi/2 \right) ] = \frac{1}{2} \left[ 1 + J_0(2 \chi )\right] +  
\sum_{i=1}^{\infty} J_{2i} (2 \chi) \cos \left(i \phi - i \pi\right) ,  \label{FATI}
\end{equation}
and
\begin{equation}
\sin ^{2} \left[ \chi \cos \left( \phi/2 \right) \right] = \frac{1}{2} \left[ 1 - J_0(2 \chi )\right] +  
\sum_{j=1}^{\infty} J_{2j} (2 \chi) \cos \left(j \phi - (j-1) \pi\right) ,  \label{FSB}
\end{equation}
for ATI [Eq. (\ref{FATI})] and sidebands [Eq. (\ref{FSB})], respectively.
The non-oscillatory ($\phi$-independent) background terms of the Fourier series
are given by $1+J_0(2 \chi)$ and $1-J_0(2 \chi)$ for the ATI peaks and sidebands, respectively.
For ATI peaks [Eq. (\ref{FATI})] the phase delays are $\delta_i = \pi$, for odd orders in $i$
and $\delta_i = 0$ for even orders in $i$, as long as its Fourier amplitude $J_{2i}(2\chi)>0$.
Correspondingly, for sidebands, the phase delay is $\delta_i = (i-1) \pi$, or equivalently $\delta_i = 0$
for odd orders in $i$ and $\delta_i = \pi$ for even orders in $i$, as long as $J_{2i}(2\chi)>0$.
However, e.g. between $\chi \simeq 2.57$ and $\chi \simeq 4.21$ the sign of the amplitude $J_2(2\chi)$ 
of the first harmonic ($j=1$), is reversed, thereby,
changing the phase delay of ATIs to $\delta_1 = 0$ and sidebands to $\delta_1 = \pi$, 
contrary to perturbation theory [Eqs. (\ref{FATI}) and (\ref{FSB})].
More generally, phase changes in the $i$th harmonic occur when $J_{2i}(2 \chi) < 0$.
The variation of the first few Fourier amplitudes as a function of $\chi$ is displayed in Fig. \ref{fourier-amplitudes}. It is obvious that the perturbative results [Eqs. (\ref{PATI}), (\ref{PSB})]
valid for small $\chi$ cease to be valid above $\chi \gtrsim 1$ when the dominance 
of the first-order amplitude is broken and, eventually, sign reversals occur for strong fields.

For completeness we note that the present semiclassical description of the interhalf-cycle interference factor
allows also for an alternative intuitive interpretation following Ref. \cite{Boll16}.
Expressing $\chi$ in terms of the quiver vectors 
$\vec{\alpha}_{\omega }=\vec{F}_{\omega }/\omega ^{2}$ and 
$\vec{\alpha}_{2\omega }=\vec{F}_{2\omega }/\left( 2\omega \right) ^{2}$
with amplitudes $F_{\omega }/\omega ^{2}$ and $F_{2\omega }/\left( 2\omega \right) ^{2}$
of the respective $\omega $ and $2\omega $ fields,
\begin{equation}
\chi = \alpha_{\omega} k_{z}\left(1+F_{2\omega }/(3\omega k_{z})\right)/\sqrt{2}
= \vec{\alpha}_{\omega }\cdot \vec{k}\left(1 + 2\omega \vec{\alpha}_{2\omega }\cdot \vec{k}/(3E_{z})\right)/\sqrt{2} ,
\label{chi}
\end{equation}
the inter-half cycle factors can be written as
\begin{eqnarray}
\cos ^{2}\left[\chi \cos \left( \phi /2\right) \right] &=&\cos ^{2}\left[ \vec{k}%
\cdot \left( \vec{R}_{+}-\vec{R}_{-}\right) /2\right] \qquad \text{\textrm{(ATI)}}
\label{molecule-ATI} \\
\sin ^{2}\left[\chi \cos \left( \phi /2\right) \right] &=&\sin ^{2}\left[ \vec{k}%
\cdot \left( \vec{R}_{+}-\vec{R}_{-}\right) /2\right] \qquad \text{\textrm{(SB)}}  
\label{molecule-SB}
\end{eqnarray}
Eqs. (\ref{molecule-ATI}) and (\ref{molecule-SB}) can be interpreted as the interference 
between the emission from two point sources located at 
$\vec{R}_{\pm}=\pm \vec{\alpha}_{\omega }\left( 1+2 \omega \vec{\alpha}_{2\omega }\cdot \vec{k}/(3E_{z})\right) \cos \left( \phi /2\right) /\sqrt{2}$. 
This picture offers a close analogy to emission from a homonuclear diatomic molecule aligned
along the polarization axis. In the case of the ATIs [Eq. (\ref{molecule-ATI})],
these two point sources emit in phase and constructive interference occurs for perpendicular emission.
For sidebands, the two point sources emit out of phase, leading
to partial destructive interference in the perpendicular direction since in
this case $\chi =F_{\omega }F_{2\omega }/(3\sqrt{2}\omega ^{3}) \neq 0$.
We want to point out that all additional approximations to SPA [analytical SPA in 
Eqs. (\ref{interf}-\ref{molecule-SB})] are performed for the sake of clarly tracing patterns in
the DDMD back to interference processes. These come at the price of limited quantitative accuracy.
In next section we explore the accuracy of the SPA.

\section{Comparison of DDMD within the SPA, SFA, and TDSE}
\label{comparison} 

\begin{figure}[tbp]
	\includegraphics[width=9cm]{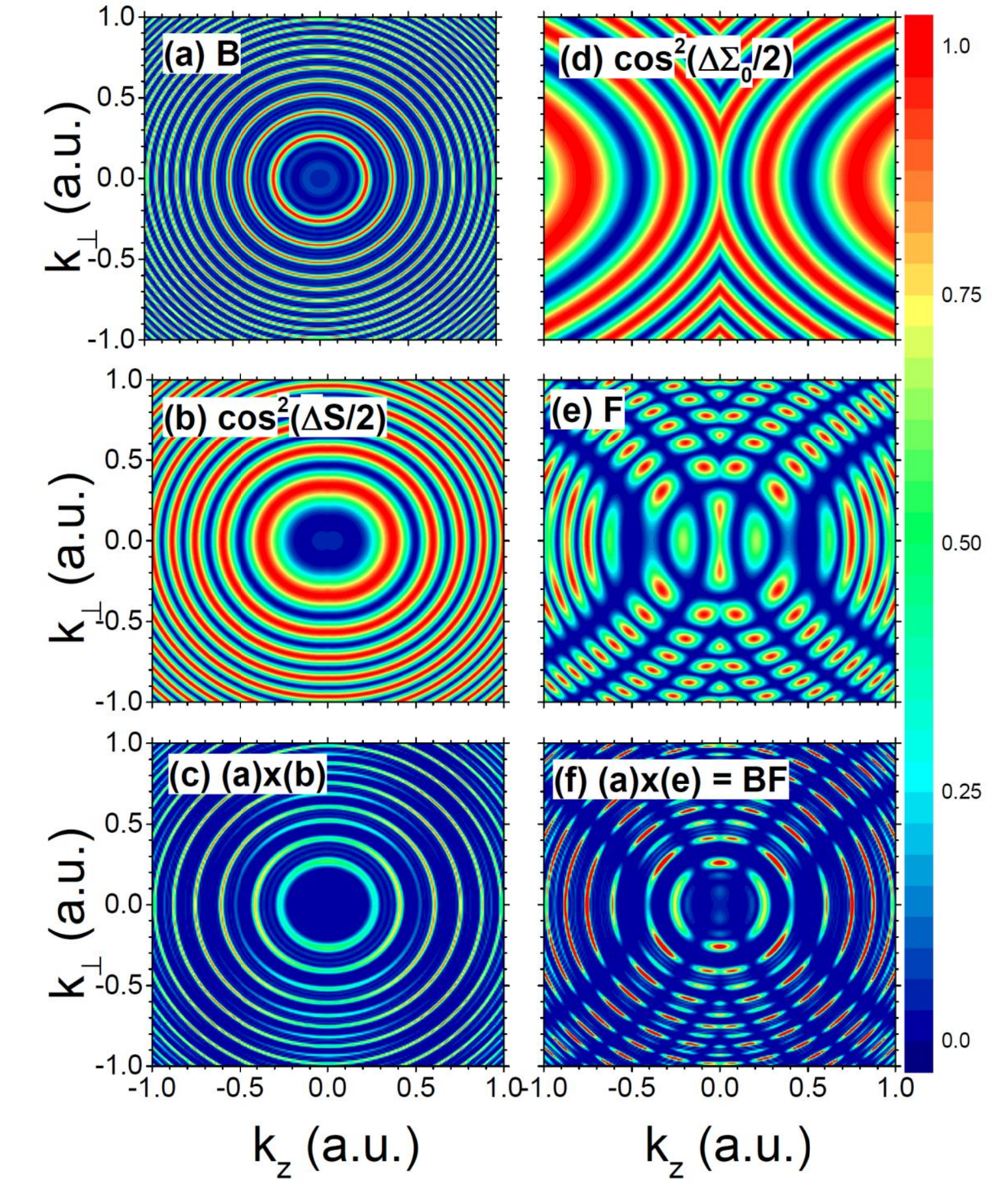}
	\caption{Doubly differential interference pattern as a function of the longitudinal $k_{z}$
	and perpendicular momenta $k_{\bot }$ for the $\omega -2\omega $ ionization
within the analytic SPA for relative phase $\phi = 3\pi/2$. (a) intercycle Bragg factor
[Eq. (\ref{inter-factor})], (b) inter-halfcycle factor [Eq. (\ref{interf})], (c) multiplication of (a) and (b), (d) intra-halfcycle factor [Eq. (\ref{interf})], (e) intracycle factor [multiplication of (b) and (d) in Eq. (\ref{factorization})], and (f) total interference pattern [multiplication of (a) and (e)]. We use a laser field with $F_{2\omega}=0.05$ a.u. $F_{2\omega}=0.005$, $\omega=0.05$, with $N=4$.}
\label{kzkrho-SPA-interf}
\end{figure}

The interplay between the Bragg (intercycle) factor [Eq. (\ref{inter-factor})] 
and the structure (intracycle) factor [Eq. (\ref{intra-factor})]
in the build-up of the interference pattern in the DDMD for ionization by the 
$\omega-2\omega$ field with $\omega = 0.05$ and
relative phase $\phi=3\pi/2$ is illustrated in Fig. \ref{kzkrho-SPA-interf} with the analytic SPA.
The Bragg factor $B(\tilde{E})$ (for $N=4$ in Fig. \ref{kzkrho-SPA-interf}a) generates
isotropic rings in the ($k_z,k_{\bot}$) plane (or spherical shells in 3D) with
radii $k_n=\sqrt{2 E_n}$ corresponding to ATI peaks (even $n$) and sideband peaks (odd $n$).
The number of (hardly seen) minima between consecutive multiphoton rings is $N-1=3$.
As expected, this factor stemming from intercycle interferences is not only independent of the emission angle but also of the relative phase $\phi$ between the $\omega$ and the $2\omega$ fields.
The inter-halfcycle factor $\cos^2(\Delta S/2)$ in Fig. \ref{kzkrho-SPA-interf}b [Eq. (\ref{interf})]
consists of a set of deformed concentric rings, slightly stretched along the longitudinal momentum and,
consequently, the isotropy is lost. 
Therefore, the minima of these rings do not perfectly match 
with those of the sidebands of the Bragg factor (Fig. \ref{kzkrho-SPA-interf}a).
When the two factors are multiplied with each other 
a modulation of these rings emerge (Fig. \ref{kzkrho-SPA-interf}c).
The intra-halfcycle factor $\cos^2(\Delta \Sigma_0/2)$ (Fig. \ref{kzkrho-SPA-interf}d) 
features an entirely different pattern of two partially overlapping deformed structures
\cite{Arbo14a,Arbo10a,Arbo12}.
The product of the inter-halfcycle (Fig. \ref{kzkrho-SPA-interf}b) and intra-halfcycle
(Fig. \ref{kzkrho-SPA-interf}d) interference factors in Eq. (\ref{interf}) approximately representing the form factor $F(\vec{k})$ is shown in Fig. \ref{kzkrho-SPA-interf}e.
The nearly isotropic inter-halfcycle factor is now strongly modulated by the intra-halfcycle pattern
resulting in a highly anisotropic emission distribution and pronounced variation with the emission angle.
Finally, the complete interference pattern in Eq. (\ref{factorization})
(Fig. \ref{kzkrho-SPA-interf}f) results from the
multiplication of the form factor $F(\vec{k})$ (Fig. \ref{kzkrho-SPA-interf}e) 
with the Bragg factor $B(\tilde{E})$ (Fig. \ref{kzkrho-SPA-interf}a).
Obviously, the resulting interference pattern is primarily governed
by that of the form factor (Fig. \ref{kzkrho-SPA-interf}e).

\begin{figure}[tbp]
	\includegraphics[width=12cm]{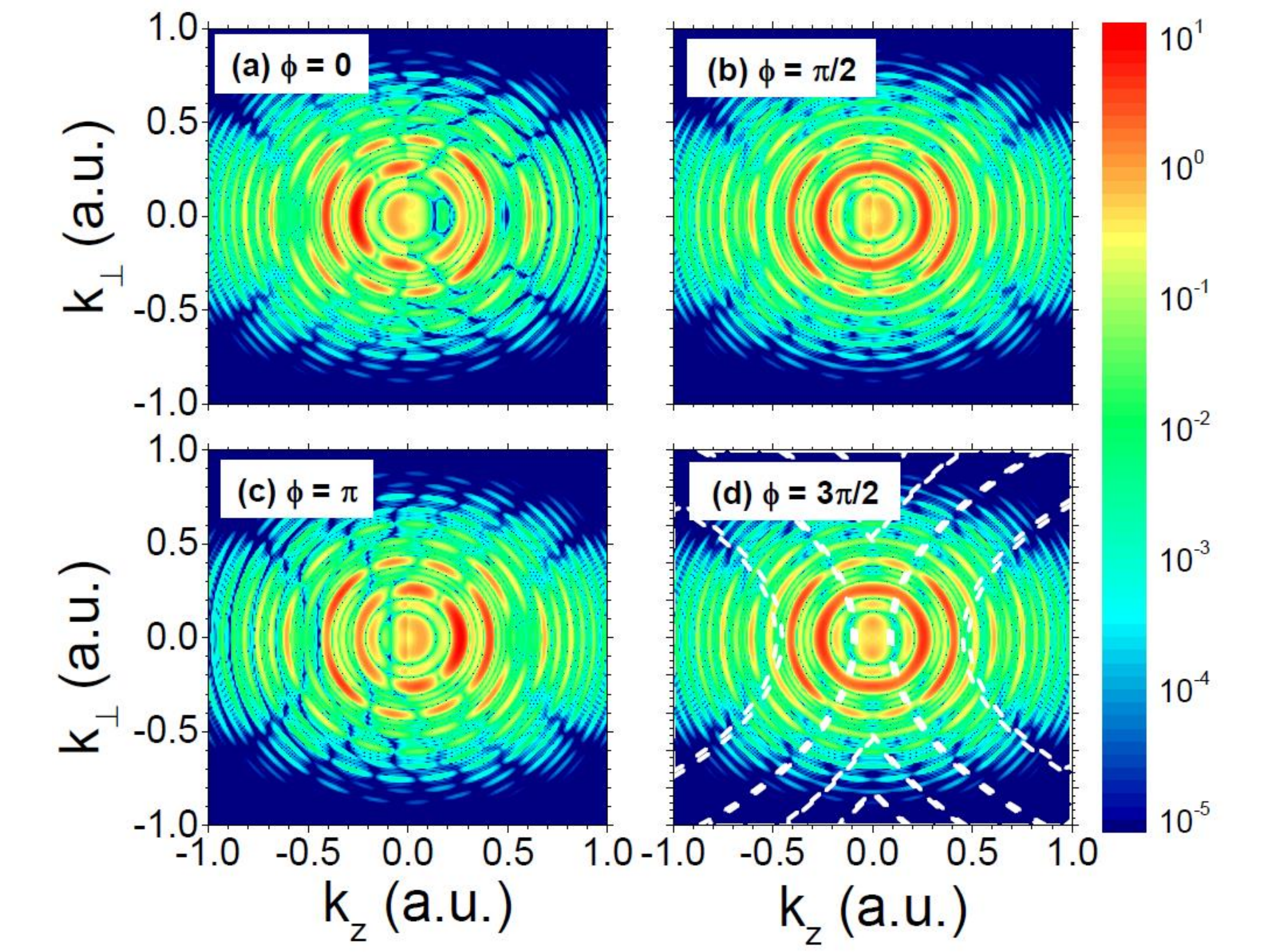}
	\caption{DDMD within the full SPA calculated using Eqs. (\ref{dPdE}) and (\ref{I1}) using the exact complex coefficients for ionization in the $\omega - 2\omega$ laser field (parameters as in Fig. \ref{kzkrho-SPA-interf}) for different phases $\phi$ between the $\omega$ and $2\omega$
	fields. (a) $\phi =0$, (b) $\phi =\pi /2$, (c) $\phi =\pi$, and (d) $\phi =3\pi /2$. In (d) we have included the nodal contours of the intra-halfcycle pattern of Fig. \ref{kzkrho-SPA-interf}d}.
\label{kzkrho-SPA-phi}
\end{figure}

The dependence of the DDMD [Eq. (\ref{dPdE})] on the relative phase within the full SPA [Eq. (\ref{I1})]
is displayed in  Fig. \ref{kzkrho-SPA-phi} for different values of the relative phase $\phi$. 
For a quantitative comparison with the SFA and TDSE results, in Fig. \ref{kzkrho-SPA-phi} we show the full SPA 
[Eq. (\ref{I1})] with complex coefficients $W(t_{\beta})$ as described in Eq. (\ref{C}) in the
appendix, without invoking the additional approximations used in arriving at Eq. (\ref{interf})
(Fig. \ref{kzkrho-SPA-interf}). While, overall, the interference pattern remains unchanged, differences in the intensity distributions in the DDMD become visible.
In general, the weak $\omega$ field (slightly) breaks the forward-backward ($k_z \longleftrightarrow -k_z$) symmetry of the DDMD within the SPA, except for $\phi=3\pi/2$, for which the vector potential remains
inversion-antisymmetric with respect to the center of the unit cell in the presence of $F_{\omega}$ and
$F_{2\omega}$ (see Fig. \ref{fields-2c}d). In Fig. \ref{kzkrho-SPA-phi}d we have also included a dashed line illustrating the contours of minima of the intra-halfcycle pattern of Fig. \ref{kzkrho-SPA-interf}. 
This structure is clearly visible in the full SPA.

\begin{figure}[tbp]
	\includegraphics[width=12cm]{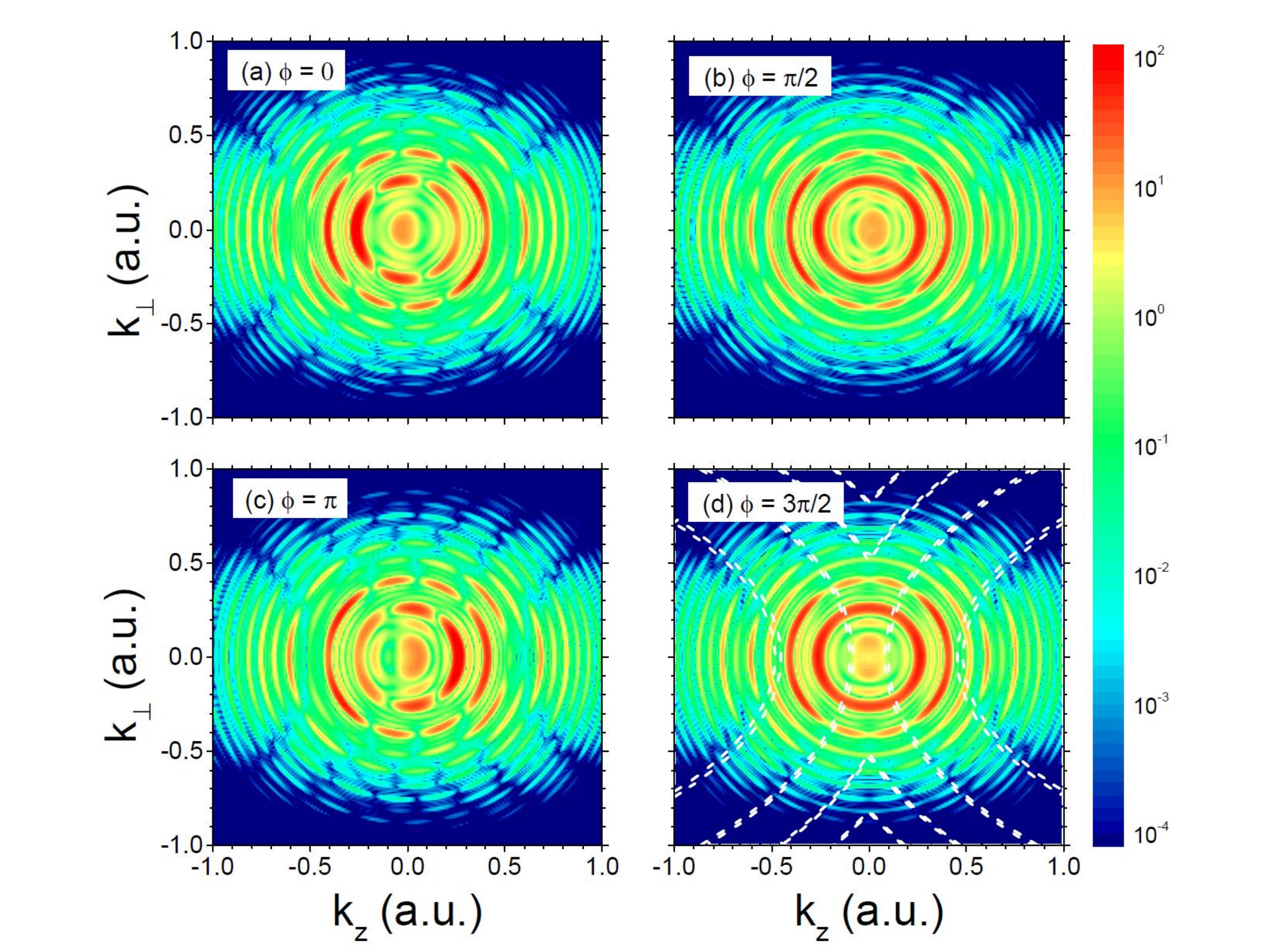}
	\caption{DDMD within the SFA for ionization in the $\omega - 2\omega$ laser field (parameters as in Fig. \ref{kzkrho-SPA-interf}) for different phases $\phi$ between the $\omega$ and $2\omega$
	fields. (a) $\phi =0$, (b) $\phi =\pi /2$, (c) $\phi =\pi$, and (d) $\phi =3\pi /2$. 
	Laser parameters as in Figs. \ref{kzkrho-SPA-interf}d and \ref{kzkrho-SPA-phi}.
	In (d) we have included the near-zero contours (for 0.01) of the intrahalf-cycle pattern of Fig. \ref{kzkrho-SPA-interf}d.}
\label{kzkrho-SFA-phi}
\end{figure}

\begin{figure}[tbp]
	\includegraphics[width=12cm]{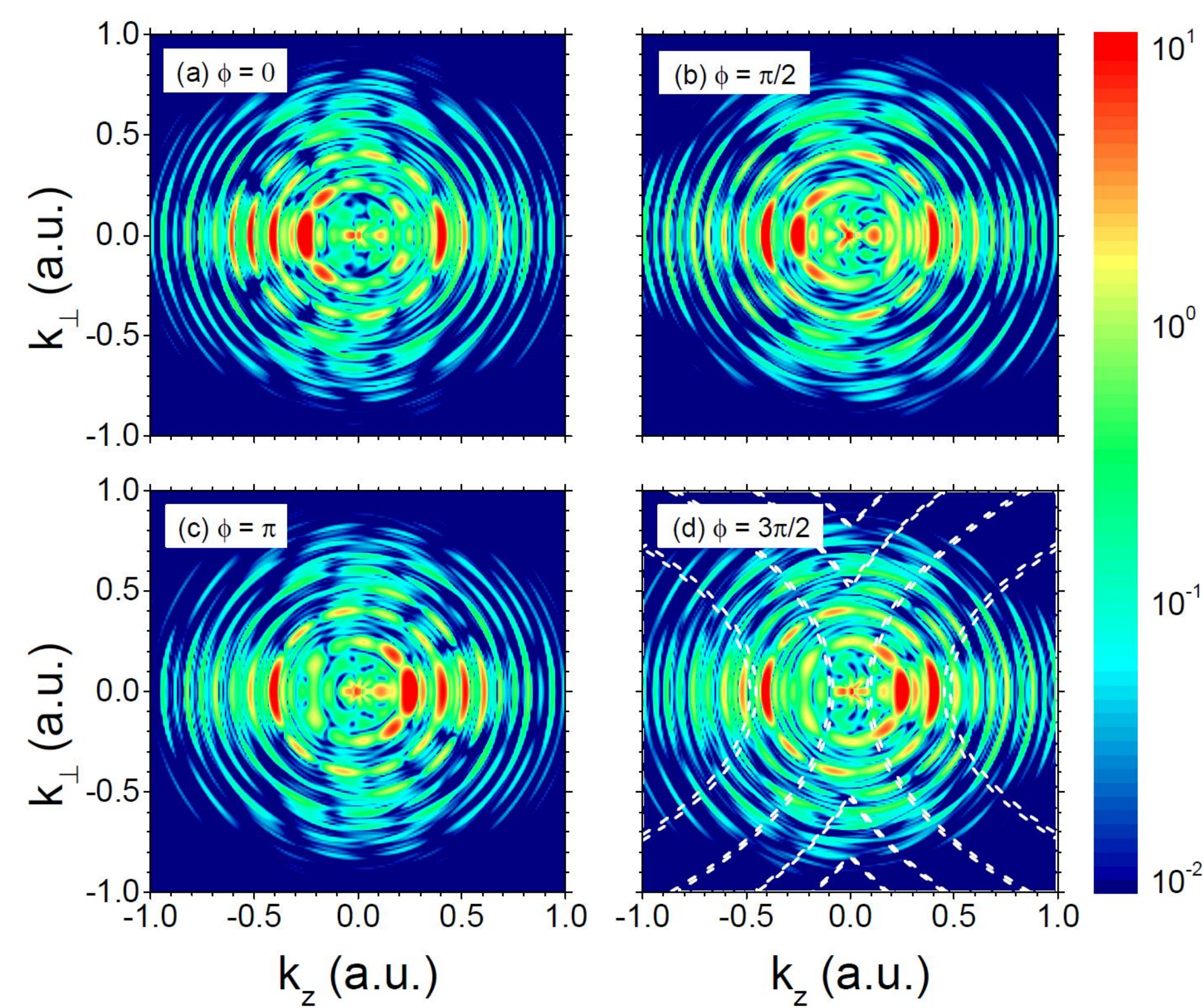}
	\caption{DDMD within the TDSE for ionization in the $\omega - 2\omega$ laser field 
	for different phases $\phi$ between the $\omega$ and $2\omega$
	fields. (a) $\phi =0$, (b) $\phi =\pi /2$, (c) $\phi =\pi$, and (d) $\phi =3\pi /2$.
	Laser parameters as in Figs. \ref{kzkrho-SPA-interf}d, \ref{kzkrho-SPA-phi}, and \ref{kzkrho-SFA-phi}.
	In (d) we have included we have included the near-zero contours (for 0.01) of the intrahalf-cycle pattern of Fig. \ref{kzkrho-SPA-interf}d.}
\label{kzkrho-TDSE-phi}
\end{figure}

We now compare the semiclassical DDMD predicted by the full SPA (Fig. \ref{kzkrho-SPA-phi}) with the
numerical evaluation of the strong-field approximation (SFA) [Eq. (\ref{Tm})]
(Fig. \ref{kzkrho-SFA-phi}) as well as with the full numerical solution of 
the time-dependent Schr\"{o}dinger equation (TDSE) (Fig. \ref{kzkrho-TDSE-phi}).
Unlike the SPA, both the SFA and the TDSE account for the envelope of the pulse
with a ramp-on and ramp-off of duration $2\pi/\omega$ each and a flat-top of duration $8\pi/\omega$.
They, furthermore, include the coupling matrix elements $M_{\mathrm{if}}$  
and, in the case of the TDSE, also the full Coulomb interaction. 
Whereas distributions for $\phi =3\pi/2 $ in SPA and SFA (Figs. \ref{kzkrho-SPA-phi}d and \ref{kzkrho-SFA-phi}d) exhibit forward-backward symmetry, the momentum distributions for other 
values of the phase, ($\phi =0, \pi /2$, and $\pi$) result in a small asymmetry
enhancing either forward or backward emission.
Pronounced forward-backward asymmetries in electron emission were experimentally observed for
$\omega-2\omega$ laser pulses for pump and probe intensities of comparable magnitude, i.e.,
$F_{2\omega}\approx F_{\omega}$ \cite{Arbo14a,Arbo15}.
The origin of this asymmetry for $\phi \neq 3\pi/2$ can be traced to the broken inversion anti-symmetry
of the vector potential relative to the center of the temporal unit cell 
(see Figs. \ref{fields-2c}a, b, and c). This fact results in differences in the actions
of the wavepackets emitted in different parts of the unit cell \cite{Arbo15}.
Note that the \textit{ab initio} solution of the TDSE in the single-active-electron approximation in the length gauge \cite{Tong97,Tong00,Arbo15} 
exhibits for the DDMD a slight forward-backward asymmetry also at $\phi =3\pi/2$ 
(Fig. \ref{kzkrho-TDSE-phi}d), however, for a different reason: here it is the effect of the Coulomb potential (absent in both the SPA and SFA calculations) on the emitted electron rather than the broken inversion anti-symmetry that causes this distortion. Comparison between Figs. \ref{kzkrho-SPA-phi}, 
\ref{kzkrho-SFA-phi}, and Fig. \ref{kzkrho-TDSE-phi} shows that the Coulomb potential also distorts the intracycle interference pattern in agreement with previous studies with one-color pulses \cite{Arbo14a,Arbo14b}. Notwithstanding the aforementioned differences,
the qualitative agreement between the TDSE, SFA, and SPA DDMD is remarkable.
Therefore, the present semiclassical model is well suited to qualitatively explain the origin
of the observed structures in terms of path interferences between
emission at different ionization times in the $\omega - 2\omega$ field.

\section{Angle-resolved energy spectra and phase delays} 
\label{delays}

\begin{figure}[tbp]
	\includegraphics[width=10cm]{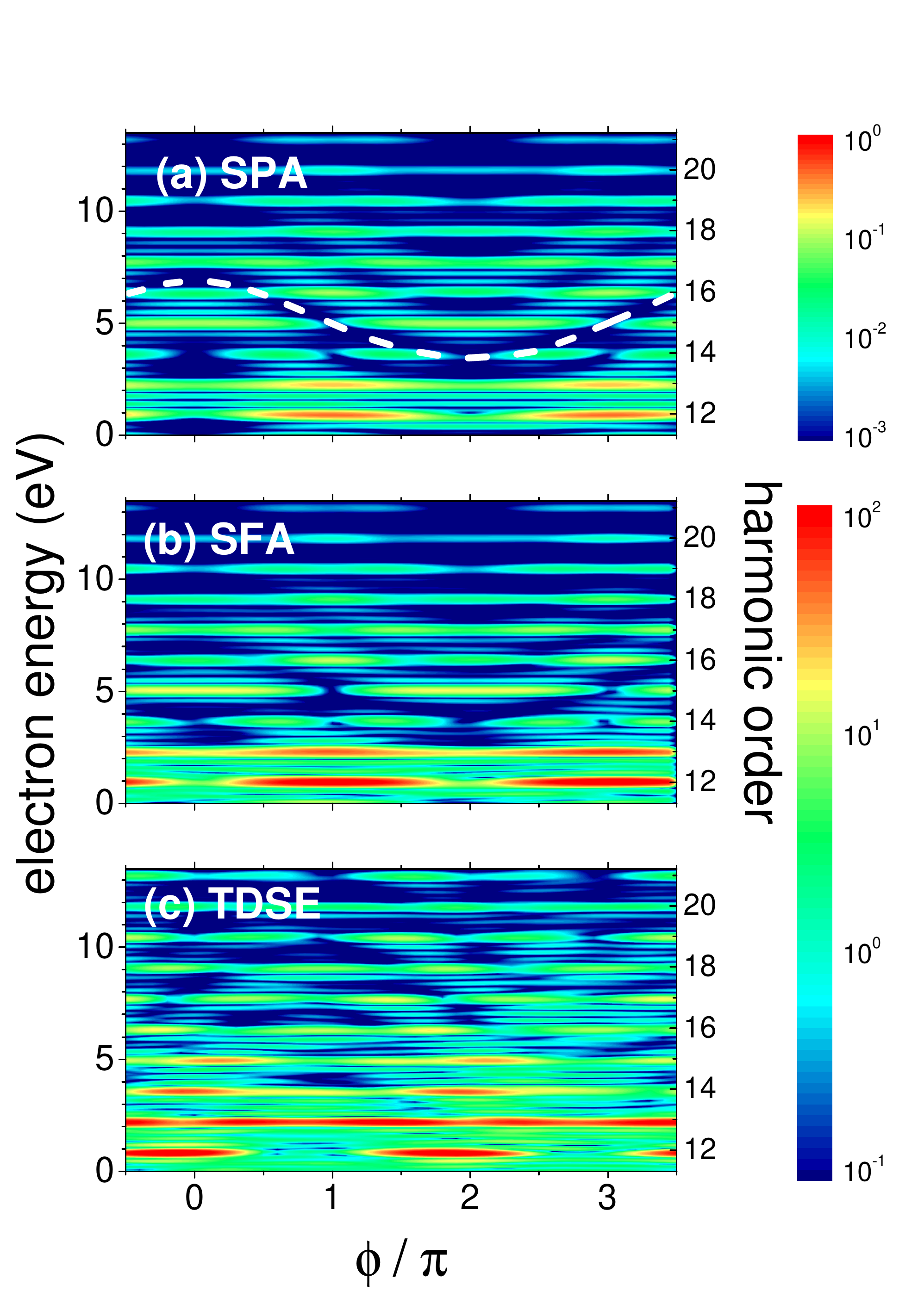}
	\caption{Differential multiphoton ionization spectrum in forward direction $P(E,\cos \theta =1)$
	as a function of the $\omega -2\omega $ phase $\phi $. (a) SPA, (b) SFA, and (c) TDSE.
	Laser parameters are the same as in 	Figs. \ref{kzkrho-SPA-interf}, \ref{kzkrho-SPA-phi}, \ref{kzkrho-SFA-phi}, and \ref{kzkrho-TDSE-phi}.
	In (a) we have included a contour of the interhalf-cycle pattern $\cos^2(\Delta S/2)$ [Eq. (\ref{interf})] to guide the eye.} 
\label{forward-phi}
\end{figure}

We finally address the angularly resolved multiphoton ionization spectrum
$P(E,\cos \theta)$ [Eq. (\ref{dPdE})], which has been recently experimentally and theoretically investigated \cite{Zipp14,Lopez21,Arbo12,Song18}.
As a prototypical case, we consider the forward emission spectrum $P(E,\cos \theta = 1)$
as a function of the relative phase $\phi$ between the two-color components with the goal to extract
the atomic multiphoton strong-field interference (MPSFI) phases $\delta_j(E)$ and their corresponding harmonic weights $c_j(E)$ [see Eq. (\ref{fourier})].

Fig. \ref{forward-phi} represents the forward spectrum calculated within the full SPA [using Eq. (\ref{I1})
without further approximations], the SFA [using Eq. (\ref{Tm})], and the TDSE
as a function of the relative phase $\phi$ between the two color components.
The ATI peaks and sidebands are formed by intercycle interferences [Eq. (\ref{factorization})] 
with peaks at energies given by Eq. (\ref{ATI1}) for $n$ even and odd, respectively.
The ATI peaks and sidebands probability densities change as the relative phase is varied.
Qualitatively, all three calculations using the SPA, the SFA, and the TDSE agree with each other.
The inter-halfcycle interference pattern  as a function of $\phi$, highly visible in the SPA multiphoton
ionization spectrum is somewhat blurred in the SFA and TDSE spectra but still present
underscoring the utility of the SPA to unravel the semiclassical origin of the $\phi$ variations.
As indicated, the inter-halfcycle interference manifests itself in Fig. \ref{forward-phi}a
as a wavy pattern as a function of $\phi$.
Changes in the multiphoton peaks and sidebands as a function of $\phi$ are primarily due to the
inter-halfcycle factor. They result from the interference between emission during the first and the second half cycles of the temporal unit cell. Each inter-halfcycle ``wavy stripe'' has
period $4\pi$ and extends from a minimum energy at $\phi=2\pi$ to a maximum energy at $\phi=0$.
For small $\chi$, the interplay of the intercycle interference given by Eq. (\ref{inter-factor})
and the inter-halfcycle interference in Eq. (\ref{interf}) causes the ATIs (sidebands) to peak at 
$\phi=\pi$ and $3\pi$ ($\phi=0$ and $2\pi$) as predicted by the perturbative limit
[Eqs. (\ref{PATI}) and (\ref{PSB})].
Increasing energy and strength of the $F_{\omega}$ probe field, and, consequently, the parameter $\chi$,
we find deviations from these perturbative predictions, and consequently, higher-order Fourier components [Eqs. (\ref{FATI}) and (\ref{FSB})] appear (see Fig. \ref{fourier-amplitudes}) and additional peaks arise.
Furthermore, higher Fourier components are responsible for the splitting of the peaks of the 
higher-lying sidebands as a function of $\phi$ (in the present case starting with the second sideband
at energies $E \gtrsim 5$eV).
For the laser parameters used in our simulations only the first (perturbative) and second Fourier components significantly contribute. Higher Fourier components will contribute for $\chi \gtrsim 2$,
or equivalently $E\gtrsim 15$ eV (above the energies considered in this paper). 
The present SPA provides useful guidance to identify the non-linear effects
in photoelectron emission for two-color fields.
Comparing the full SPA, the SFA, and the TDSE, the discrepancies among the three calculations,
in particular near threshold, can be traced back primarily to two features: (i) the SPA fails to
fully reproduce the SFA because the action $S(t)$ is not large and, therefore, the semiclassical limit
is not reached and (ii) the SFA fails to accurately reproduce the TDSE since the influence of the Coulomb potential is still significant.

\begin{figure}[tbp]
	\includegraphics[width=10cm]{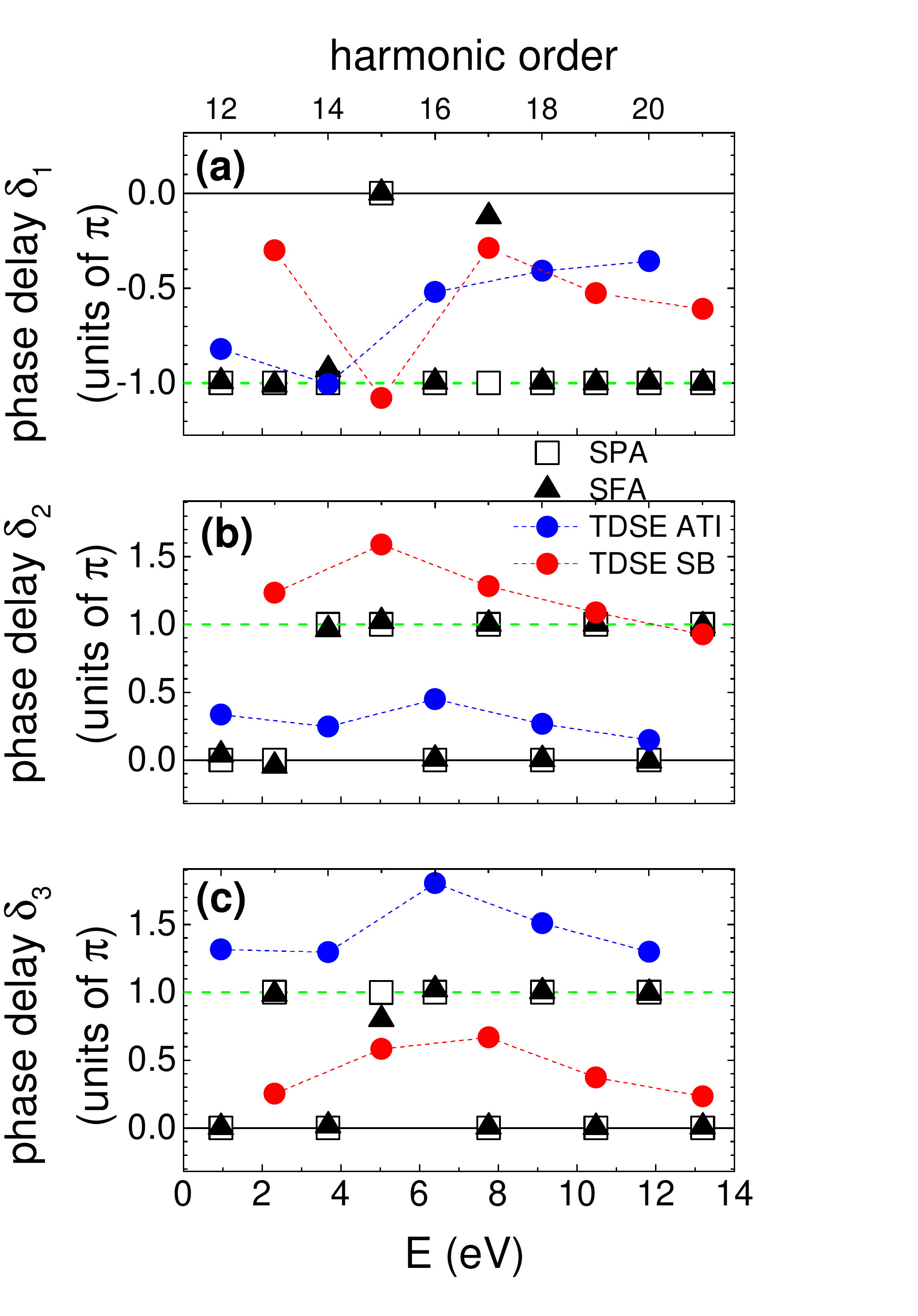}
	\caption{MPSFI phases $\delta_1(E_n)$ (a), $\delta_2(E_n)$ (b), and $\delta_3(E_n)$ (c) [Eq. (\ref{fourier})] 	extracted from the spectra within the SPA, SFA, and TDSE in Fig. \ref{forward-phi} at ATI and sideband energies given by Eq. (\ref{ATI1}).}
\label{fourier-delays}
\end{figure}

A Fourier analysis of the angular differential emission spectrum in Fig. \ref{forward-phi} 
[Eq. (\ref{fourier})] allows the extraction of the set of energy dependent atomic phase delays
[$\delta_i(E)$, $i=1,2,\ldots$]. In the present case contributions from the first three harmonics
$i=1,2,3$ can be clearly identified (Fig. \ref{fourier-delays}) indicating that interferences
between quantum paths differing by up to $3$ ($2\omega$)-photons (Fig. \ref{paths}a-c)
effectively contribute. Correspondingly, up to $n_{\omega}=6$ $\omega$-photons are involved to close the interference loops (see Fig. \ref{paths}).
This illustrates the mayor structural difference between the present MPSFI and the standard RABBIT
protocol. Within the SPA and SFA, the phase shifts $\delta_1(E)$, $\delta_2(E)$, and $\delta_3(E)$ are
found to be (mostly) either $0$ or $\pi$ and agree with each other, with the notable exception
$\delta_1(E_{17})$ for harmonic energy $E_{17}=7.75$eV to be discussed below. This overall agreement
illustrates the applicability of the SPA to estimate atomic SFA phase delays.
However, significant departures from $0$ or $\pi$ delays arise for the TDSE
due to the effect of the Coulomb potential of the ionic core on the outgoing electron.
These deviations are more pronounced for the first Fourier component $\delta_1$.
In turn, for the second and third orders, $\delta_2(E)$ and $\delta_3(E)$ appear to converge
to the SFA predictions as $E$ increases.

\begin{figure}[tbp]
	\includegraphics[width=10cm]{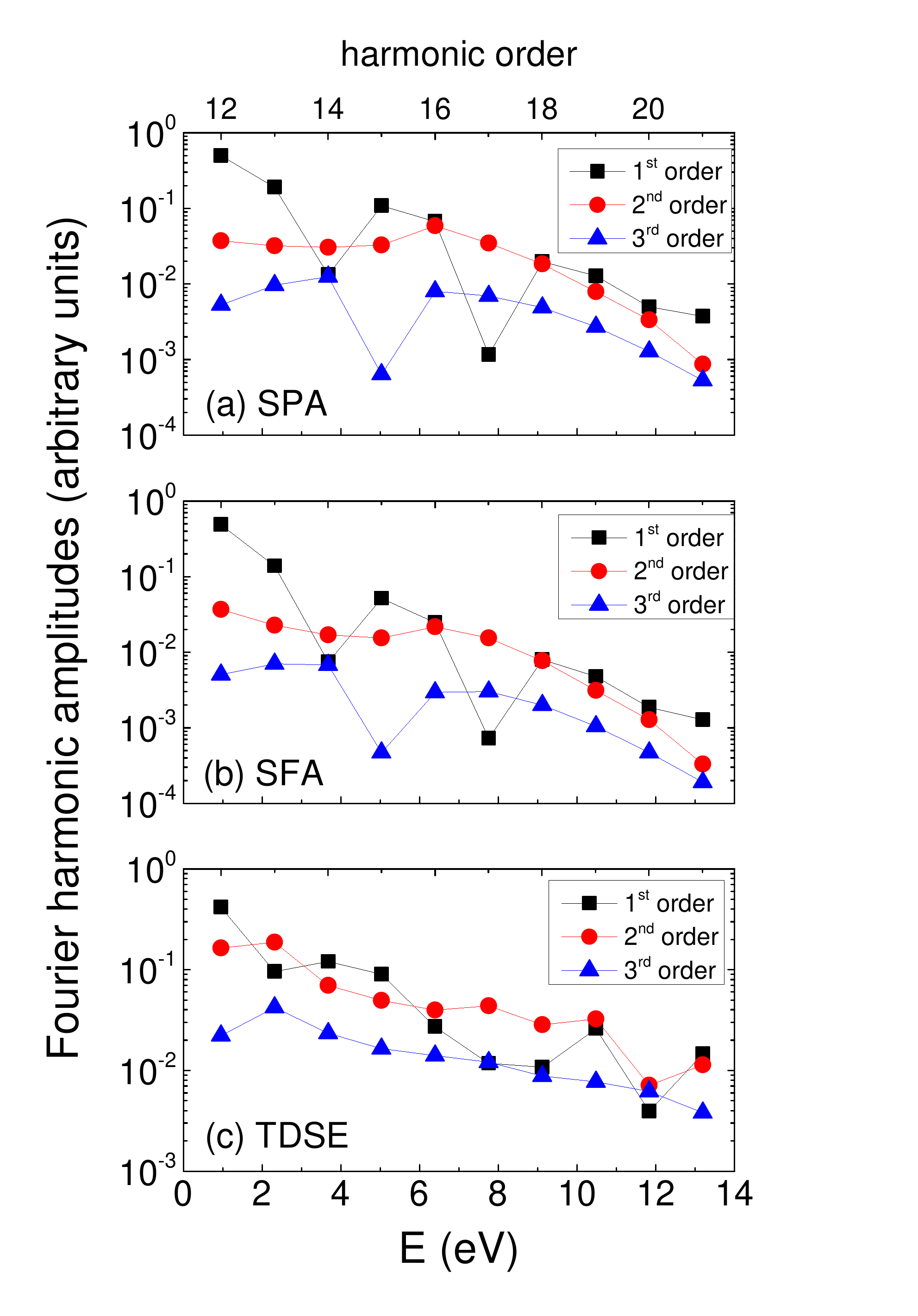}
	\caption{Harmonic amplitudes $c_1(E_n)$ (a), $c_2(E_n)$ (b), and $c_3(E_n)$ (c) [Eq. (\ref{fourier})] extracted from the spectra of Fig. \ref{forward-phi} as a function of $E_n$ calculated with the SPA, the SFA, and the TDSE.}
\label{harmonic-amplitudes}
\end{figure}

Turning now to the amplitudes $c_j(E_n)$ of these Fourier components, we observe in 
Fig. \ref{harmonic-amplitudes}, overall, a decrease of the amplitudes with increasing energies $E_n$
for both the ATI peaks ($n$ even) and sidebands ($n$ odd). A clear dominance of the first harmonic component
which involves paths that differ by only
one strong-field $2\omega$-photon can be found only near threshold. 
With increasing energy the relative importance of higher Fourier components grows.
In particular, near the sideband $n=17$, the amplitude $c_1(E_{17})$ is very small compared to 
$c_2(E_{17})$ and $c_3(E_{17})$ in the SPA and SFA, and to $c_2(E_{17})$ for TDSE.
At this energy the first-harmonic order amplitude is about one order of magnitude smaller than the corresponding third-order amplitude and two orders of magnitude smaller than the corresponding second-order amplitude. Therefore, the mismatch between the SPA and SFA phases $\delta_1(E_{17})$ mentioned above 
(Fig. \ref{fourier-delays}) is likely caused by the numerical uncertainty. 

Overall, the Fourier expansion [Eq. (\ref{fourier})] features pronounced high-order Fourier components thereby precluding the characterization of the MPSFI signal by a single energy dependent phase $\delta_1(E)$. Consequently, also the direct association with a time delay related to the spectral derivative of this phase, $d\delta/dE$, is no longer possible. Characterization of the emitted wavepacket in a (non-perturbative) strong-field $\omega - 2\omega$ setting requires, in general, a multitude of phases. Only in the perturbative limit of a weak $\omega$ field, the reduction to a single phase in analogy to the RABBIT-like protocol is justified. The present analysis provides the underlying background and conceptual insights into the recent numerical observation \cite{Lopez21} that the atomic phase delay $\delta_1(E)$ features a sensitive dependence on the probe field, contrary to the expectations suggested by the perturbative limit.

\section{Conclusions}
\label{conclusions}

We have developed a semiclassical non-perturbative strong field theory
for the atomic ionization by a linearly polarized $\omega -2\omega $ laser pulse. 
While the $\omega$-probe field is assumed to be weaker than the strong $2\omega$ pump field,
($F_{\omega} < F_{2\omega}$), its amplitude $F_{\omega}$, as employed in recent experiments \cite{Zipp14},
is found to be sufficiently strong as to open a plethora of interfering quantum pathways of absorption and
emission of $2\omega$ and $\omega$ photons. The interferences resulting from this multi-photon
strong-field ionization process need to be characterized by an entire set of atomic phase delays
$\left\{\delta_i(E)\right\}$ with $i=1,\ldots$ rather than a single phase delay $\delta(E)=\delta_1(E)$,
as customary in a structurally similar perturbative RABBIT-like setting. 
Each phase $\delta_i(E)$ can be associated with a class of pairs of
interfering pathways differing by the absorption of $i = \left| n_{2\omega} - n^{\prime}_{2\omega} \right|$ $2\omega$ photons. The present semiclassical theory allows mapping these phases onto temporal
interferences between wavepackets emitted during different cycles of the $2\omega$ field denoted as
inter-halfcycle interferences. These emissions may occur within the same optical cycle of the $\omega$
field which defines the length of the temporal unit cell.
We present the angular differential emission spectrum in three different approximations: the
semiclassical stationary phase approximation (SPA), the strong-field approximation (SFA), and the
numerical solution of the time-dependent Schr\"{o}dinger equation (TDSE).
We find that phase delays calculated within the SPA agree not only with the present SFA
calculations but also with previous perturbation theories \cite{Zipp14, Lopez21} when the probe field amplitude is much weaker than the pump pulse, i.e., $F_{\omega} \ll F_{2\omega}$.
The extension to stronger fields gives rise to novel effects at intermediate and high electron energies of forward spectra such as the departure from the sinusoidal modulation of the intensity of ATI peaks and
sidebands as a function of the relative phase $\phi$ between the two-color components of the electric field.
This results in the splitting of the  maxima of ATIs and sidebands as a function of the two-color phase $\phi$. Phase shifts $\delta_j$ associated with higher-order Fourier components, with $i \geq 2$,
are a signature of the non-linearity in the probe field. For the parameter lasers used in the
experiment, i.e., $F_{2\omega}/F_{\omega}=0.1$, the second harmonic of the phase delay significantly
contributes to the spectrum. Therefore, the strong influence of non-linear contributions on electron photoemission in pump-probe setup points to the need to revise the extraction method of phase delays beyond the perturbative regime.


\medskip

\section*{Acknowledgements}

This work was supported by CONICET PIP0386, PICT-2017-2945, PICT-2020-01434,
and PICT-2020-01755 of ANPCyT (Argentina) and by the Austrian FWF (grant Nos. M2692, W1243).
D.G.A especially thanks S. Eckart, M. Dahlstr\"{o}m, and M. Bertolino for fruitful discussions.

\appendix*
\section{Saddle-point integration}

In this appendix we calculate the time integral appearing in the
transition amplitude stemming from a single optical cycle, $I_{\mathrm{if}}(\vec{k})$,
[Eq. (\ref{intra-I})] by means of the saddle point approximation. We evaluate
\begin{equation}
I_{\mathrm{if}}(\vec{k})=-i\int_{0}^{T}\vec{F}(t)\cdot \vec{d}\left[ \vec{k}+%
\vec{A}(t)\right] e^{iS(t)}dt  ,
\label{intra-I-appendix}
\end{equation}
where the dipole transition moment is defined as $\vec{d}(\vec{v})=(2\pi
)^{-3/2}\langle e^{i\vec{v}\cdot \vec{r}}|\vec{r}|\varphi _{i}(\vec{r}%
)\rangle $, and the phase in Eq. (\ref{intra-I-appendix}) is given by the Volkov action in Eq. (\ref{action}) 
\cite{Volkov35}. The dipole matrix elecment coupling a $1s$ hydrogenic initial state to a
final Volkov state can be calculated as
\begin{equation*}
\vec{d}(\vec{v})=\frac{1}{(2\pi )^{3/2}}\int d\vec{r}e^{i\vec{v}\cdot \vec{r}%
}\ \vec{r}\ \varphi _{1s}(\vec{r}),
\end{equation*}%
with $\varphi _{1s}(\vec{r})=(Z^{3/2}/\sqrt{\pi })\exp (-Z/r),$ where $Z^{2}/2=I_{p}$.
This integral is given in terms of the momentum representation of the hydrogenic wave function by
\begin{equation*}
\vec{d}(\vec{v})=\frac{\sqrt{2}(2I_{p})^{5/4}}{\pi }\frac{i\vec{v}}{\left(
v^{2}/2+I_{p}\right) ^{3}} .
\end{equation*}
Consequently, Eq. (\ref{intra-I-appendix}) can be written as
\begin{eqnarray}
I_{\mathrm{if}}(\vec{k}) &=&\frac{\sqrt{2}(2I_{p})^{5/4}}{\pi }\int_{0}^{T}%
\frac{\vec{F}(t)\cdot \left[ \vec{k}+\vec{A}(t)\right] }{\left( \frac{\left[ 
\vec{k}+\vec{A}(t)\right] ^{2}}{2}+I_{p}\right) ^{3}}e^{iS(t)}dt  \notag \\
I_{\mathrm{if}}(\vec{k}) &=&-\frac{\sqrt{2}(2I_{p})^{5/4}}{\pi }\int_{0}^{T}%
\frac{\ddot{S}(t)}{\left[ \dot{S}(t)\right] ^{3}}e^{iS(t)}dt, 
\label{Iif-app}
\end{eqnarray}
where $\ddot{S}(t)=-\vec{F}(t)\cdot \left[ \vec{k}+\vec{A}(t)\right] $ and $\dot{S}(t)$
is given by Eq. (\ref{action}). Since the zeros in the denominator of the integrand of 
Eq. (\ref{Iif-app}) coincide with the saddle point condition $\dot{S}(t_{\beta })=0$ 
[Eq. (\ref{saddle})], the standard saddle point approximation cannot be directly applied.
Instead, expanding the denominator to first order around the saddle point
\begin{equation*}
\dot{S}(t) \simeq \dot{S}(t_{\beta })+\ddot{S}(t_{\beta })(t-t_{\beta }).
\end{equation*}
leads to
\begin{equation}
I_{\mathrm{if}}(\vec{k})=-\sum_{t_{\beta }}\frac{\sqrt{2}(2I_{p})^{5/4}}{\pi %
\left[ \ddot{S}(t_{\beta })\right] ^{2}}\int_{0}^{T}\frac{e^{iS(t)}}{%
(t-t_{\beta })^{3}}dt.  
\label{Iif3}
\end{equation}
Eq. (\ref{Iif3}) features a third-order singularity and thus can be analytically evaluated following 
Eq. (B6) of Ref. \cite{Gribakin97} to yield
\begin{equation}
I_{\mathrm{if}}(\vec{k})\simeq \sum_{\beta }W(t_{\beta })e^{iS(t_{\beta })}.
\end{equation}
with complex coefficients
\begin{equation}
W(t_{\beta })\simeq -\frac{2\sqrt{2}(2I_{p})^{5/4}e^{i\alpha (t_{\beta })}}{%
\left\vert F(t_{\beta })\right\vert \sqrt{2I_{p}+k_{\perp }^{2}}},
\label{C}
\end{equation}
and phases $\alpha(t{\beta }) =-\arg \ddot{S}(t{\beta })$.

\bibliographystyle{unsrt}
\bibliography{biblio-diego}

\end{document}